\documentclass[11pt,onecolumn,a4paper]{article}
\usepackage[latin1]{inputenc}
\usepackage{epsfig}
\usepackage[english]{babel}
\usepackage{textcomp}
\usepackage{amsmath}
\usepackage{amsfonts}
\usepackage{amssymb}
\usepackage{float}
\usepackage{xcolor}
\usepackage{abstract}
\usepackage{array,multirow}
\usepackage{graphicx}
\usepackage{subcaption}
\usepackage{authblk}
\usepackage{color, colortbl}
\usepackage[autostyle]{csquotes}

\title{\bfseries Electron Beam Chirp Dexterity in Staging Laser Wakefield Acceleration}

\author{$\rm N. ~Pathak^{1,2}$\thanks{naveenpathak@sanken.osaka-u.ac.jp}}
\author{$\rm A. ~Zhidkov^{1,2}$}
\author{$\rm T. ~Hosokai^{1,2}$}

\vspace{5.0cm}

\affil{$\rm ^{1}$ \small {\em Institute of Scientific and Industrial Research (ISIR), Osaka University, Mihogaoka 8-1, Ibaraki, Osaka, 567-0047, Japan.}}
\affil{$\rm ^{2}$\small {\em Laser Accelerator R$\&$D , Innovative Light Sources Division, RIKEN SPring-8 Center, 1-1-1, Kouto, Sayo-cho, Sayo-gun, Hyogo, 679-5148, Japan}}

\date{}

\begin{document} 
\maketitle

\begin{abstract}
Particle energy chirp is shown to be a useful instrument in the staging laser wake field acceleration directed to generation of high-quality dense electron beams. The chirp is a necessary tool to compensate non-uniformity of acceleration field in longitudinal direction and achieve essential reduction of energy dispersion. This is demonstrated via particle-in-cell simulations exploiting the splitting technique for plasma and beam electrons.  Properly chosen beam chirps allow decrease in the energy dispersion of order of magnitude in every single stage during acceleration to the GeV energy range.
\end{abstract}

\newpage

\section*{I. Introduction}
Laser plasma accelerators (LPA) have made outstanding progress in the last two decades \cite{Gibbon,DUmstadter,Joshi,Esarey1,VMalka,SMHooker} . A major impetus in this field comes from the development of petawatt class laser pulses and their quasi controlled propagation in specially designed plasma channels \cite{Gonsalves}. As a result, energy gain of the accelerated electron beams in multi-GeV range \cite{Gonsalves,WPLeeman2006,Wang2013, HTKim} has been demonstrated in few centimeters single stage laser wakefield acceleration (LWFA). Simultaneously, significant progress has also been made in achieving unprecedented ultra-short \cite{Lundh}, low emittance \cite{Barber,Qin,Plateau}, high pointing stability \cite{Hosokai} and novel steering of the electron beams \cite{Nakanii} from LWFA. However, in spite of these outstanding progresses, use of LWFA accelerated electron beams for potential scientific and industrial applications are limited due to their relatively large energy spreads. For example, LWFA based extremely compact source of multi-GeV electron beams for free-electron lasers (FELs) require below percent level energy spread. Therefore, mitigating the problem of the large relative energy spread is one of the prime concerns in LWFA research. 

In general, the problem of energy spread in LWFA is indispensable and associated with the three fundamental reasons: (i) an uncontrolled or continuous self-injection of the electrons in the laser wakefield, (ii) charge density of the injected electron beam beyond the beam loading limitation, and (iii) the non-uniform profile of the wakefield strength in the longitudinal direction. In the uncontrolled or continuous self-injection, the injected electrons accumulate a random phase-space, and consequently, acquire a broad energy spectrum. Several methods, viz. colliding laser pulses \cite{Umstadter,Esarey}, density gradient injection \cite{SBulanov,Suk,Tomassini,Ohkubo}, shock assisted self-injection of the electrons \cite{Schmid,Buck}, ionization injection \cite{Chen,Pak,Zhidkov}, application of an external magnetic field \cite{JVieira} and frequency chirp \cite{VBPathak,NPathak} have been proposed and implemented successfully in the experiments \cite{Faure,Geddes,McGuffey,THosokai2,Mirzaie,HTKim2} to control the self-injection of the electrons. Although, these approaches were instrumental in enhancing the reproducibility and stability of the acceleration process, they were limited in providing the improvements in the energy spread of the electron beam, which remains well above a few percent level or even more. Partially, the inadequacy of some of these processes to minimize the energy spread could be due to the strict requirements for the control of not only to localize the injection process but also to the number of injected electrons, and all with micron level precision. Using colliding laser pulse scheme Rechatin $\it{et}$ $\it{al.}$, \cite{Rechatin} has shown that the beam loading process (or charge density of the injected electrons) can be controlled and relative energy spread of $1\% $ level can be achieved. Colliding laser pulse scheme is attractive but experimentally quite challenging and require extremely stable laser system for precise spatio-temporal overlapping of the colliding pulses. Even a small longitudinal or/and transverse shot-to-shot fluctuation in the laser focal spot may not yield the desired outcome. The minimal loading of the wakefield with a tailored electron bunch in the bubble regime was theoretically investigated by Tzoufras $\it{et}$ $\it{al.}$ \cite{Tzoufras}, however, beam shaping is an another challenging crucial issue in LWFA experiments. A possible solution to these problems is to completely separate the LWFA based injection and acceleration stages. This require an extremely stable and reproducible injector with energy slicing scheme to produce electron beams with unprecedented quality, which could be injected and synchronized with the plasma wave in the booster stages. Such stable injector along with stable synchronization with the booster stage are developed for multi-stage LWFA and demonstrated experimentally \cite{Steinke,Sakai,Jin}. Alternatively, using a well-defined externally injected electron beam, for example from a conventional source, may completely avoid the problems that could arise from the self-injection process. However, the problem of energy spread still persists. If the longitudinal extent of an externally injected electron beam is not short enough than the acceleration phase of the plasma wave, which is not uniform and of the order of few tens of micrometers, then it will acquire energy spread in proportional to the field strength profile as follows. Being oscillatory in nature, the plasma wave has periodically varying acceleration (or deceleration) phase along its wavelength. For an under-loaded LWFA, in the accelerating (or decelerating) phase the profile of the longitudinal electric field can be approximated as: $E_{x2}=E_{x1}-\alpha x$ ($E_{x2}=E_{x1}+\alpha x$), where $x$ is the longitudinal position along the acceleration phase of the wakefield and $\alpha$ is the rate of change of the wakefield strength. This is illustrated by PIC simulation in fig.1(a) and fig.1(b) using the splitting technique for plasma and externally injected beam electrons. This implies that in the acceleration (or deceleration) phase the longitudinal field strength has a linear chirp. If the longitudinal extent of the externally injected (or self-injected) electron beam is comparable to the scale length of the acceleration phase ($\lambda_{p}/2$) of the plasma wave, then this linear chirp will be imprinted on the energy spectrum of the electron beam during the acceleration process: the rear part of the electron beam experiences higher acceleration field, whereas, the front part of the electron beam experiences relatively lower acceleration field [cf. fig.1(b)]. As a result, an initially un-chirped electron beam will develop a positive chirp during the acceleration process, i.e., the higher energetic particles are at the rear and the lower energetic particles are at the front of the beam. This phenomenon is illustrated numerically in fig.2. This self-induced dynamic energy chirp by the plasma wave results in a relatively large energy spread of the accelerated electron beam. 

To overcome this problem optimum beam loading effect has been proposed in order to obtain a near uniform profile of the wakefield. This can be done either by increasing the charge of the injected electron beam or by injecting another secondary auxiliary electron beam \cite{Manahan}. However, beam loading effect may not maintain the flat profile of the wakefield for a longer propagation length due to the dynamic evolution of the electron beam and the laser pulse, whereas injecting an auxiliary electron beam in not feasible experimentally for LWFA, instead it increases the complexity of the experimental approach. Furthermore, partial self-correction of the electron beam energy spread by self-excited plasma wave, exploiting plasma wave de-chirping ability, has been demonstrated experimentally for electron beams from conventional particle accelerators \cite{Shpakov}. However, such schemes may not be sufficient for electron beams generated from LWFA where energy spread is in the range of several percents or even more. 

A major breakthrough came in recently with an elegant solution to this problem, where it is shown that if an electron beam with an initially negative energy chirp will be injected in the wakefield, the effect of the self-induced energy chirp in LWFA could be potentially mitigated \cite{Pousa}. In order to simply understand the effectiveness of a negatively energy chirped electron beam in reducing the energy spread, we adopted a one dimensional heuristic approach excluding the effect of electron beam loading and evolution of the laser pulse or plasma field. One dimensional analysis are quite useful to underlying the acceleration process in nonlinear plasma waves \cite{EEsarey}.  Let's consider the equations of motion for beam electrons in a non-uniform field in the form of a running wave in 1D approximation:

\begin{equation}
E(x)=E_{max}\left(1-\frac{x}{L}+\frac{\beta t}{t_{0}}\right)^{n}=E_{max} \ \xi \nonumber
\end{equation}
where, $\xi=\left(1-\frac{x}{L}+\frac{\beta t}{t_{0}}\right)^{n}$. If $\xi$ is smaller than 0 or larger than 1 the field strength become zero. Here, $E_{max}$ is the maximal value of the field strength, $\beta=v_{gr}/c$ ($v_{gr}$ is the velocity of the wave and $c$ is the speed of light), L is the wave size, $t_{0}=L/c$ and $x$ is the coordinate. Using dimensionless variables such as $\tau=t/t_{0}$, $A=-eE_{max}L/mc^2$, $u=p/mc$, $y=x/L$, equations can be rewritten as

\begin{align}
\frac{\partial u}{\partial \tau} & = A\eta \xi^n \\ \nonumber
\frac{\partial \xi}{\partial \tau} & = \beta-\frac{u} {\sqrt{1+u^2}} \ \ ; if \ 0<\xi<1
\end{align}

we use the definition: $v/c=u/\sqrt{1+u^2}$ in 1D approximation. Solutions for $u(t)$ and $x(t)$ are complicated and weakly analyzable. Therefore, we will determine a parametric solution for $u$ as function of $u(\xi)$. $\eta$ is equal to zero if $\xi$ is out of the unit interval [0,1]. It is apparent that the acceleration is started for a particle with $\xi \sim 1$ and finished at $\xi=0$ that allows us to evaluate the energy distribution after the acceleration is completed. Using the apparent condition: $\frac{\partial u}{\partial \tau}=\frac{\partial u}{\partial \xi} \frac{\partial \xi}{\partial \tau}$ one can get the equation in the form 

\begin{equation}
\frac{\partial u}{\partial \xi}\left(\beta -\frac{u}{\sqrt{1+u^2}} \right)=A\eta \xi^n
\end{equation}

A solution for $k-th$ particle with initial conditions $u(\xi_{0k})=u_{0k}$, where $\xi_{0k}$ is determined by an initial position of $k-th$ particle, can be easily found from Eq.(2):

\begin{equation}
\beta\left[u_{k}(\xi)-u_{0k}\right]+\sqrt{1+u_{ok}^{2}}-\sqrt{1+u_{k}(\xi)^2} = A\left[\xi^{n+1}-\xi_{0k}^{n+1} \right]/(n+1)  \nonumber
\end{equation}

The acceleration stops when $\xi=0$, therefore the maximal momentum for particles is

\begin{equation}
u_{kmax}=\frac{\beta C_{k}}{1-\beta^2}+\sqrt{\beta^2 C_{k}^{2}/(1-\beta^2)^2 + C_{k}^2 -1} \approx 2C_{k}/(1-\beta^2) \nonumber
\end{equation}

where $C_{k}=\sqrt{1+u_{0k}^2}-\beta u_{0k}+A\xi_{0k}^{n+1}/(n+1)$. The best effect of the chirp can be reached if all $u_{kmax}$ are equal in spite of different $\xi_{0k}$. The energy spread in this case vanishes. Choosing a particle with position $\xi_{0M}=1$ and the initial momentum $u_{0M}$, one can get the optimal beam chirp for the linear field strength of the running wave as 

\begin{equation}
x_{k}/L=\left(u_{0k}-u_{0M}\right)\left(1-\beta^2\right)\left(n+1\right)/2A
\end{equation}

For spatial distribution according to Eq.(3) [the negative chirp] all electrons will have the same energy at the exit from acceleration phase of the field. Even for a distribution close to the ideal an essential correction in the energy dispersion can be achieved. 

The use of negative chirp, which naturally occurs during propagation of the electron beam in vacuum, is an important step towards generating an ultra-low energy spread electron beam from a laser or plasma wakefield accelerator. However, it is noteworthy that when an externally injected electron beam, with the negative energy chirp, interacts with the plasma wave its phase-space is gradually rotated by the quasi-linear positive chirp of the wakefield. Therefore, the effect of the self-induced dynamic chirp on the energy spread cannot be compensated completely for long plasma length. How long the effect of the negative energy chirp of the electron beam will be maintained is depends on the strength of the plasma wave and the initial mean energy of the injected electron beam. This implies that after a certain acceleration length the effect of the negative chirp of the electron beam will gradually begin to saturate, and once again the plasma wave induced positive chirp will starts dominating. Therefore, understanding and characterization of the dynamic evolution of the energy chirp of the electron beam interacting with the plasma wave is essential. 

In this paper, we present a comprehensive numerical investigation of dynamic evolution of the energy chirp of an externally injected electron beam with the plasma wave in LWFA configuration via self-consistent and relativistic \textit{ab-initio} Particle-In-Cell (PIC) simulations. The influence of the initial electron beam energy on the evolution of the longitudinal phase space (energy chirp) is investigated in order to understand the advantage of this scheme for both the lower as well as higher initial energy of the injected electron beam. This may partially help in the design requirements, when it is required to extend from the two-stage to multi-stage LWFA configuration. Furthermore, the role of longitudinal plasma density tailoring is shown to be instrumental in further narrowing down of the energy spread. Applicability of such schemes for an electron beam with relatively large initial energy spread is also investigated. In the present analysis, we avoid the conditions which may result in the self-injection of the electrons that lead to dark current. This may have an additional implications on the parameters (length, energy spread etc.) of the final electron beam.

\section*{II. Simulation parameters}

We conducted a numerical study of the dynamic energy chirp evolution of an externally injected electron beam with the LWFA based booster stage in two-dimensional (2D) geometry using fully relativistic PIC code FPLaser \cite{Zhidkov2}. Necessity of very high accuracy in calculations of the velocity of laser pulses in plasma and long, centimeter scale, pulse progation lengths determines our choice of the simulation geometry. The simulations were performed using moving window technique. In the simulations, the laser pulse propagates along the negative x-direction [from the right to the left side]. The parameters of the laser pulse, plasma density and injected electron beam were chosen considering the requirements for a single-stage (or booster-stage) LWFA with an external injection to emulate the two-stage acceleration \cite{Pathak}. 

The dynamic energy chirp evolution of an externally injected electron beam is studied in a parabolic plasma channel. The channel is assumed to be fully ionized for helium gas specie. Along the propagation direction, the plasma density has a linearly increasing profile up to 100 $\mu m$ and is then constant. Along the transverse direction, the plasma density has a parabolic channel to provide guiding to the laser pulse. The profile of the channel is given as $N_{e}(r)=N_{emin}+\triangle N_{e}(r/R)^{2}$, where $R=40 \ \rm \mu m$ is the radius of the channel, $N_{emin}=7 \times 10^{17} \rm cm^{-3}$ is the on-axis plasma density, and $\triangle N_{e}=0.7$ is the density depth. The effect of longitudinal density profile is also investigated. 

The driver laser pulse parameters in the booster stage were those for a standard Ti:Sapphire based laser system, which has a laser pulse of central wavelength $\lambda=800 \ nm$. The pulse duration is matched with the plasma density and its full width half maximum (FWHM) is $\tau=70 \rm \ fs$. The laser pulse energy is $E_{L}=10 \ \rm J \ (141 \rm \ TW)$ with a normalized vector potential of $a_{0}=3$ and the laser spot size is $2w_{0}=30 \ \mu m$. The parameters of an externally injected electron beam are chosen based on the matching criteria with the booster stage \cite{Pathak}. The mean energy of the electron beam is $E_{b}=100 \rm \ MeV$ with an energy spread of $\triangle E/E=3\%$. In order to study the effect of the initial mean energy on the dynamic evolution of the energy chirp, we also used an electron beam with  $E_{b}=500 \ \rm MeV$. The effect of initial energy spread of the electron beam is also investigated.  The spatial and temporal profile of the electron beam is Gaussian with diameter $2r_{b}=10 \ \mu m$  and length $L_{b}=10 \ \mu m$. The delay between the booster stage laser pulse and an injected electron beam is $100 \ \rm fs$ in order to overlap the electron beam with the acceleration phase of the wakefield in the first plasma bucket. The optimum charge of the electron beam is $10 \ \rm pC$ in order to achieve an initial quasi-uniform acceleration field of the excited wake wave. The size of the simulation window is $x \times y =80 \mu m \times 300 \mu m$. The axial resolution is $\lambda/120$ and the transverse resolution is $\lambda/32$ and there were 25 particles per cell. High spatial and temporal resolution has been chosen to diminish the effects of numerical dispersion. For detailed description of the optimization of the matching condition, such as synchronization, emittance growth, energy selection criteria and overlapping of the injected electron beam and booster stage the reader may refer to \cite{Pathak}.

\section*{III. Results and discussion: $\newline$ (a) Dynamic energy chirp evolution of the electron beam in plasma waves}
Simulations were performed for both un-chirped and chirped electron beams in the staged LWFA configuration. Figure 3(a) and 3(b) shows the initial spatial distribution for an un-chirp ($E=E_{0}-\alpha x, \alpha=0$) and negatively chirped ($E=E_{0}-\alpha x, \alpha \neq 0$) electron beams, respectively. Here $E_{0}$ is the mean energy of the electron beam, $\alpha$ is the chirp parameter and $x$ is the longitudinal position of the beam particles. Except chirp all other parameters of the two electron beams are the same (as described in Section II). In the simulations these electron beams are injected externally, with a suitable delay, in order to co-propagate with the acceleration phase of the wakefield excited by the laser pulse in the booster stage. The dynamic evolution of the longitudinal phase-space and energy spectrum in the two cases is then followed at regular intervals. Figure 4(a1-f1) and 5(g1-l1) shows the dynamic evolution of the energy chirp in the longitudinal phase-space of the electron beams for an unchirped electron beam, whereas, figure 4(a2-f2) and 5(g2-l2) shows the dynamic evolution of the energy chirp for a negatively chirped electron beam.  The rotation of the phase-space can be observed from these figures, which is induced by the chirp in the longitudinal electric field of the wakefield. Figure 6(a1-f1) and 7(g1-l1) shows the evolution of the corresponding energy spectrum for an initially unchirped electron beam, whereas, figure 6(a2-f2) and 7(g2-l2) shows for an initially negative energy chirped electron beam. One can notice the lower FWHM energy spread for an initially negative energy chirped electron beam. Figures 4(a1,2,3)-4(f1,2,3) corresponds to the simulation time of 110 fs, 550 fs, 1.0 ps, 1.3 ps, 1.6 ps, and 2.0 ps, respectively. Figure 5(g1,2,3)-5(l1,2,3) corresponds to simulation time of 2.8 ps, 5.6 ps, 22.6 ps, 23.0 ps, 23.5 ps, and 24.0 ps, respectively. For a longer simulation time or propagation length, the energy spread gradually increases even for a negatively chirped electron beam as well. Although it is not as large as in the case of unchirped electron beam, it may become above the one percent level as shown in Fig. 6(l2). 

The rotation of the phase-space takes place in about 2.8 ps time. This implies that the electron beam with an initial mean energy of 100 MeV and energy spread of $3\%$ require around 1 mm propagation length in the chosen laser and plasma parameters for complete rotation of the chirp. Up to this point the maximum compression in the energy spectrum of the electron beam can be achieved along with the energy gain. Beyond this length standard acceleration process will gradually start dominating, which will induce energy spread, however, for rest of the acceleration length this will be quite smaller as compared to the energy spread for an un-chirped electron beam. Further, it was also observed that the energy gain of the electron beam is independent of the nature of the chirp, i.e., both the un-chirp and negatively chirped electron beam experiences same amount of energy gain. These are important observations from the simulation results, which implies that manipulating the initial parameters of the electron beam may delay the phase inversion time while continuing the energy gain. This knowledge will become important when proper designing of the plasma target is required depending on the different initial mean energy of the injected electron beam or when more than two stage LWFA based acceleration process will be required to gain energies in multi GeV scale. The dependence of the initial electron beam energy and energy spread is later investigated in this section. 

\section*{(b) Effect of longitudinal plasma density tailoring}

Further improvement in the energy spectrum of the electron beam can be achieved by tailoring the longitudinal density profile of the plasma channel. By longitudinal tailoring of the plasma density the electron beam can be re-phased with the laser wakefield \cite{Guillaume,ADopp}. Moreover, density tailoring will change the de-phasing length over which the electron beam runs out of phase with the accelerating part of the plasma wave and enter in the decelerating part. Energy spread reduction by increasing longitudinal plasma density or decreasing dephasing length is useful in the case of self-injection schemes where the electrons with higher energy are at the front and the lower energy one are at the rear part of the beam \cite{Dopp}. Contrary to this scenario, for an externally injected un-chirped or negatively chirped electron beam the energy chirp induced by the plasma wave in the booster stage is exactly opposite, i.e., higher energy electrons are at the rear and lower energetic electrons are at the front of the beam. Thus, the principle adopted in \cite{Dopp} for reducing the energy spread is not directly applicable for an externally injected electron beam. However, it is important to take into account that by increasing the plasma density the plasma length decreases, which means the wakefield slips forward with respect to the electron beam. Now, If the re-phasing of the electron beam can be achieved in a manner that the higher energy electrons at the back of the beam lose energy, whereas, the lower energy electron at the front of the beam gain energy, then it allows narrowing down the energy spectrum of the electron beam. In experiments, one of the simplest ways of achieving a reproducible longitudinal density tailoring is tilted gas-jet or specially designed gas-jets \cite{Jinnew}. Such schemes can be extended to tapered capillary discharge or gas-cell. In simulations, we used a linearly increasing density profile along the longitudinal or laser propagation direction. The linear density profile is applied after 6 mm propagation length, where the energy spread of the negatively chirped electron beam increases beyond the one percent level. Figure 8 illustrate the 1D profile of the plasma density. The density is increased by a factor of two within a scale-length of 1 mm.  By increasing the plasma density, the wavelength of the wakefield decreases. As a result, the electron beam gradually slips out of phase with the plasma wave: the plasma wave tend to move forward with respect to the electron beam. Now, a part of the rear side of the electron beam (higher energy component) enter partially in the decelerating phase of the wakefield, whereas, the leading part of the electron beam (lower energy component) enter the relatively higher electric field. Moreover, the electric field strength of the wakefield also increases slightly due to the increase in the plasma density. The higher electric field provide relatively higher energy gain. Figure 4(a3-f3) and 5(g3-l3) shows the dynamic evolution of the energy chirp in the longitudinal phase-space of the electron beams, whereas, fig. 6(a3-f3) and 7(g3-l3) shows the evolution of the corresponding energy spectrum for an initially negatively chirped electron beam with longitudinal plasma density tailoring. It is clear from these simulation results that the role of plasma density tailoring is extremely useful is achieving an ultralow energy spread. The rate of change (or increment) of the plasma density and the corresponding scale-length are required to be optimized with the longitudinal extent of the electron beam to achieve the lowest possible energy spread. 

\section*{(c) Effect of initial mean energy of the injected electron beam}

Next, we examine the effect of the \enquote{initial mean energy} of the electron beam on the dynamic chirp rotation by the plasma wave. The initial energy of the electron beam is increased from 100 MeV to 500 MeV. All other parameters remained unchanged. Figure 9 and 10 shows the dynamic energy chirp evolution for this case. Figures 9(a), 9(b), 9(c), 9(d), 9(e) and 9(f) corresponds to the simulation time of 1.2 ps, 2.7 ps, 5.5 ps, 8.8 ps, 10.5 ps, and 13.0 ps, respectively. Figures 10(a), 10(b), 10(c), 10(d), 10(e) and 10(f) corresponds to the simulation time of 15.0 ps, 17.3 ps, 19.3 ps, 21.5 ps, 22.5 ps, and 24.0 ps, respectively. Unlike to the lower initial mean energy case, now the plasma wave requires longer time to rotate the chirp from negative to positive. In the case of 100 MeV electron beam the chirp of the electron beam is rotated in about 2.8 ps, whereas, in the case of 500 MeV electron beam it takes $\sim$17 ps. This implies that higher energetic electron beam will require longer plasma length for rotation of the initial negative chirp. This may be advantageous because in longer plasma length the energy gain of the electron beam is also higher (provided the pump depletion length and dephasing lengths are sufficiently long). As shown in fig. 10(f) the energy gain after 7.0 mm is similar as in the case of 100 MeV electron beam.  As found in the simulations, the energy spread after 7.0 mm propagation length is also similar to 100 MeV case $(0.45\%)$. Smaller energy spread of $\sim 0.36\%$ can be seen at slightly lower energy gain (fig. 10(c-e)). Simulation results suggest that higher initial electron beam may be attractive to achieve lower energy spread even at higher energy gain. At this point one can immediately conclude that in future, employing several LWFA based multiple acceleration stages, along with chirp rotation technique, may allow higher energy gain as well as state-of-the-art lower energy spread, which is necessary for potential practical applications. In the ongoing quest for the LWFA seeded compact X-FEL, it is possible to achieve below $1\%$ or lower energy spread at few-GeV class energy through staging, however, this needs to be optimized for the desired set of the experimental conditions. 

\section*{(d) Effect of initial energy spread of the injected electron beam}

Finally, we examine the effect of the \enquote{initial energy spread} of the injected electron beam on the evolution of the dynamic chirp. This is useful to understand if a negatively chirped electron beam with an arbitrary energy spread can be applied or not.  The initial energy of the electron beam is 100 MeV. The energy spread is now increased from $3\%$ to $10\%$. All other parameters remained unchanged. Figure 11 shows the evolution of the longitudinal phase space and energy spectrum of the electron beam in this case. Figure 11(a) and 11(b) corresponds to the simulation time of 110 fs and 23.0 ps, respectively. The energy spectrum shows double peak structure. The FWHM spread of each peak is $\sim 5.3$ and $5.1 \ \rm MeV$. This implies energy spread of $\sim 1\%$ in each peak. Although simulation result does not show a single peak, we anticipate that by proper optimization of the laser pulse and the plasma density parameters single peak can be obtained and even a larger energy spread beam can be used in staging LWFA experiments employing the negatively chirped electron beam. This may increase the flexibility of operation with LWFA based injector stage. 

The simulation results are summarized in table 1. At this point we would also like to clarify that the evolution of the laser pulse and hence the wakefield is quantitatively differ in 2D and 3D simulations, however, qualitatively the result will not differ significantly. We anticipate that for the same amount of beam charge one may expect relatively higher energy gain and narrow energy spread in 3D simulations as compared to the 2D simulation results. This is because the laser pulse intensity evolves as $w_{0}^2$ in 3D, whereas it evolves as $w_{0}$ in 2D. Here $2w_{0}$ is the full waist size of the focus spot. Following Ref. \cite{Pathak}, higher intensity laser pulse in the booster stage may compensate better the effect of the beam loading of the injected electron bunch.

\section*{IV Conclusions}
The laser wakefield acceleration is produced by a longitudinally non-uniform electric field. For reasonable beam charges different parts of electron beam are exerted by different acceleration force. Even an initially mono-energetic electron beam will acquire large energy dispersion in longer acceleration runs. Since this is the principle characteristic of laser pulse wake one has only one way: manipulation by the parameters of the injected electron beam. 

To find processes resulting in a reduction of energy spread of accelerated electrons we have carried out 2D particle-in-cell simulations splitting plasma electrons and electrons constituting injected beam. Electron beams with an initially unchirped and negatively chirped energy spectrum are used for the comparative study. Calculations have been performed from injected electron beams with an initial energy of 100 MeV and 500 MeV and charge 10 pC in plasma with its length about cm scale and irradiated by sub-PW class laser pulses. Electron acceleration process is analyzed for different initial energy spread of the electron beam and plasma density profile.

The use of the negative chirp for electron beams in the staging laser wake field acceleration has been shown as a flexible procedure allowing control with an essential reduction of the beam energy dispersion in every acceleration stage. Negatively chirped electron beams with an initial mean energy of 100 MeV and energy spread of $3 \%$ can be reduced to $\sim 0.46 \%$ energy spread in 7 mm propagation length and with an energy gain of $>$ 400 MeV. Moreover, it was shown that even a higher initial energy spread of $10 \%$ can be reduced down to $1 \%$, although independently in two multiple peaks. We anticipate that with a more careful tuning of the parameters, electron beam with single peak and energy spread under $1 \%$ can be obtained for initially higher energy spread beam as well. For an initial higher mean energy of 500 MeV and energy spread of $3 \%$, the reduction in the energy spread is shown to be $\sim 0.45 \%$ in 7 mm propagation length after the acceleration up to GeV level. Such dispersion correction has been demonstrated only for one stage. It is apparent that further improvement can be achieved in the next stage(s) for the beam having already higher energy and lower energy dispersion and eventually, energy spread below $0.1\%$ could be achieved in multi-stage LWFA configuration. 

The longitudinal tailoring of the plasma density is shown to be an  instrumental parameter for the reduction of the beam energy spread. Electron beams can be re-phased with the laser wakefield. This gives further reduction of the energy dispersion compare to the uniform plasma. The dispersion correction with the negative chirp depends on beam energy (or/and energy spread) as seen from Eq. (3) and supported by our numerical simulations. 

The problem of necessary chirp formation can be partially solved by using magnetic chicane \cite{Reiser}. For higher energetic electron beams ($>$ few hundreds of MeV) one can use magnetic chicane for energy dispersion manipulation. However, for relatively smaller energetic electron beams ($\leq$ 100 MeV) the negative chirp can be easily formed in vacuum for a quite clear reason \cite{Jin}. A parameter determines the chirp characteristic here is the propagation length. However, as shown in \cite{Jin}, the use of low density plasma with the beam self-focusing effect may essentially improve the chirp technique resulting in almost mono-energetic distribution of accelerated electrons suitable for X-FEL applications.

\section*{Acknowledgement} 
This work was funded by the JST-MIRAI program grant no. JPMJMI17A1, and was partially supported by the ImPACT R$\&$D Program of Council for Science, Technology and Innovation (Cabinet Office, Government of Japan). We are grateful to Prof. Yuji Sano, and Dr. Kando (QST-KPSI) for their encouragements and helpful discussions. We also acknowledge the use of Mini-K computing facility at SACLA, RIKEN, SPring-8 Center. 

{\bf Note: The data that support the findings of this study are available from the corresponding author upon reasonable request.}



\newpage

\begin{table}
\begin{center}
\begin{tabular}{|c|c|c|c|c|c|c|c|}\hline
\multirow{2}{2.0cm}{Beam charge    (pC)}  & $\rm E_{bi}$ &  \centering $ \rm \triangle E_{bi}$ & \centering Beam  & Plasma & $\rm E_{bf}$ & $\rm \triangle E_{bf}$ \\  & \centering (MeV) & \centering (fwhm, $\%$) & chirp & profile & \centering (MeV) &  (fwhm, $\%$)\\ \hline
\multirow{1}{*}{10} 
    & 100 & 3 & 0 & uniform & $\sim 500$ & $\sim 4$ \\ \hline  
\multirow{1}{*}{10} 
    & 100 & 3 & -ve & uniform & $\sim 500$ & $\sim 1.1$ \\ \hline       
    \multirow{1}{*}{10} 
    & 100 & 3 & -ve & with up-ramp & $\sim 570$ & $\sim 0.46$ \\ \hline
    \multirow{1}{*}{10} 
    & 100 & 10 & -ve & with up-ramp& $\sim 485$ & double peak \\ \hline
    \multirow{1}{*}{10} 
    & 500 & 3 & -ve & with up-ramp & $\sim 915$ & $\sim 0.45$ \\ \hline  
\end{tabular}
\end{center}
\caption{Summary of the acceleration of an externally injected $10 \ \rm pC$ electron beam in the booster stage with different plasma density profile and electron beam initial parameters. Here, $\rm E_{bi}$ is the initial mean energy of the externally injected electron beam, $\rm \triangle E_{bi}$ is the initial energy spread, $\rm E_{bf}$ is the final energy and $\rm \triangle E_{bf}$ is the final energy spread of the electron beam after acceleration in the booster stage.}
\end{table}

\newpage


\begin{figure}
\centering
\includegraphics[width=1.0\linewidth]{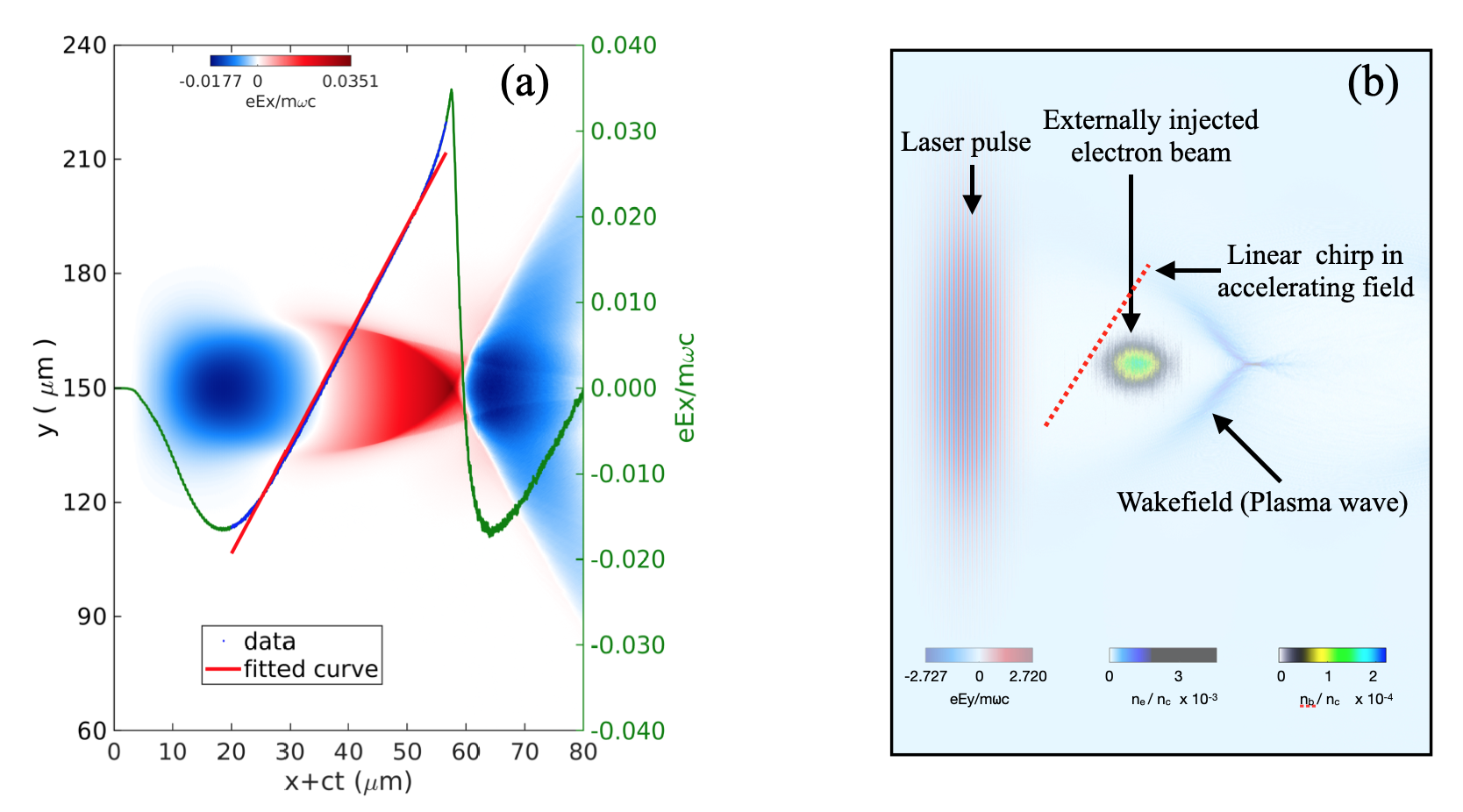}
\caption{\small PIC simulation results: (a) illustrating the evolution of the normalized longitudinal electric field ($eE_{x}/m\omega c$) of the plasma wave. The laser pulse is propagating from right to left side. The blue region of the colorbar represents the decelerating field, whereas, red region is the accelerating field. The 1D line plot in green color is the on-axis accelerating field. The 1D blue curve is the data set used for curve fitting and the red color 1D line is the fitted curve. The fitted curve representing the linear chirp in the longitudinal electric field of the plasma wave. (b) illustrating the evolution of the laser wakefield and trapping of an externally injected electron beam in the acceleration phase of the first plasma bucket. The red dotted line represent the linear chirp of the accelerating field over the entire electron beam. The color bar represent normalized laser pulse transverse electric field field ($eE_{y}/m\omega c$), normalized plasma electron density ($n_{e}/n_{c}$) and normalized externally injected electron beam density ($n_{b}/n_{c}$).}\label{fig:1}
\end{figure}


\begin{figure}
\centering
\includegraphics[width=1.0\linewidth]{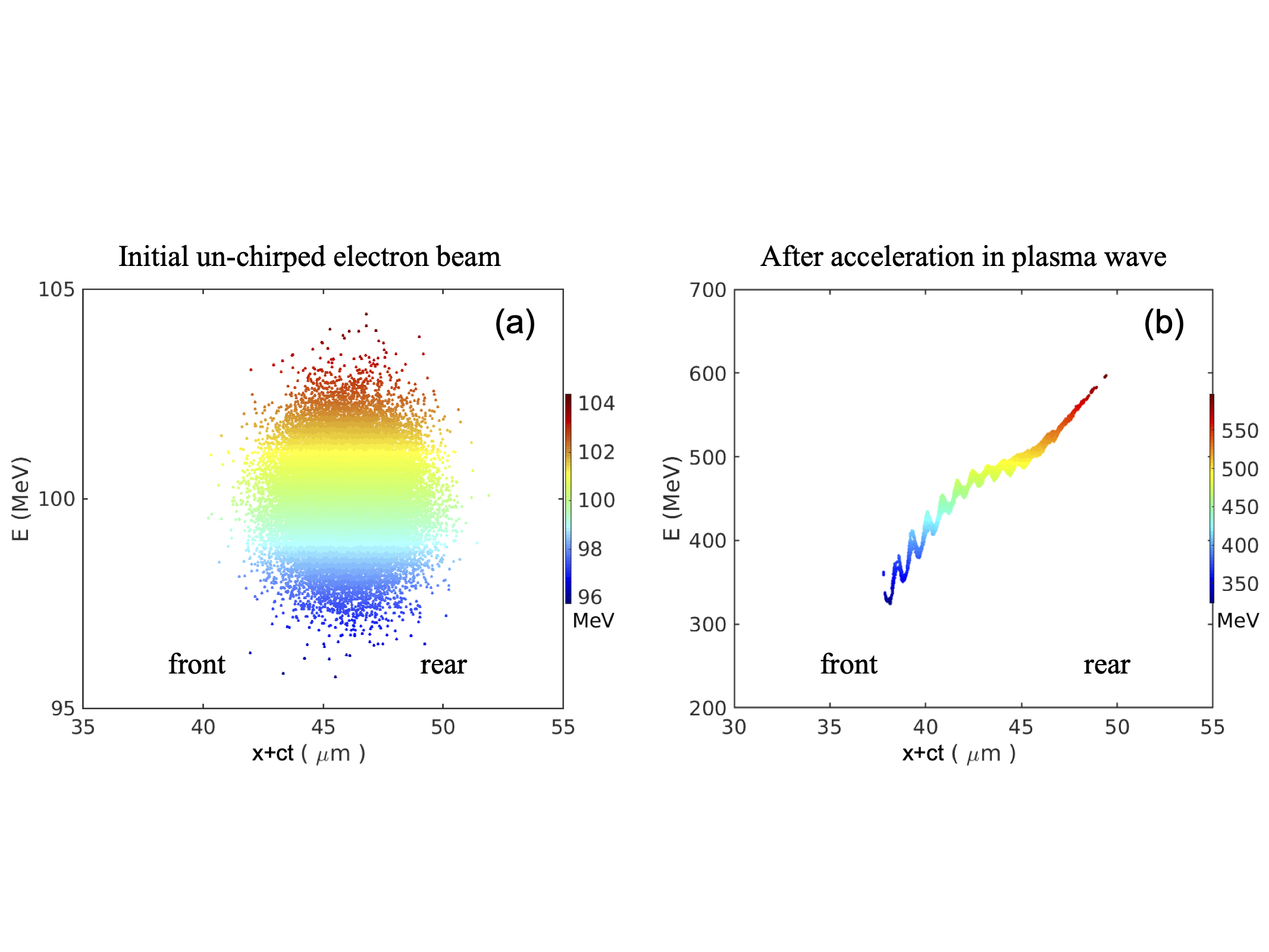}
\caption{\small (a) Longitudinal phase-space of an unchirped electron beam and (b) is the longitudinal phase-space of the electron beam after interacting with the acceleration phase of the plasma wave. It shows the linear chirp of the longitudinal electric field of the plasma wave is imprinted in the phase-space of the accelerating electron beam. The electron beam is propagating from right to left side.}\label{fig:2}
\end{figure}


\begin{figure}
\centering
\includegraphics[width=\linewidth]{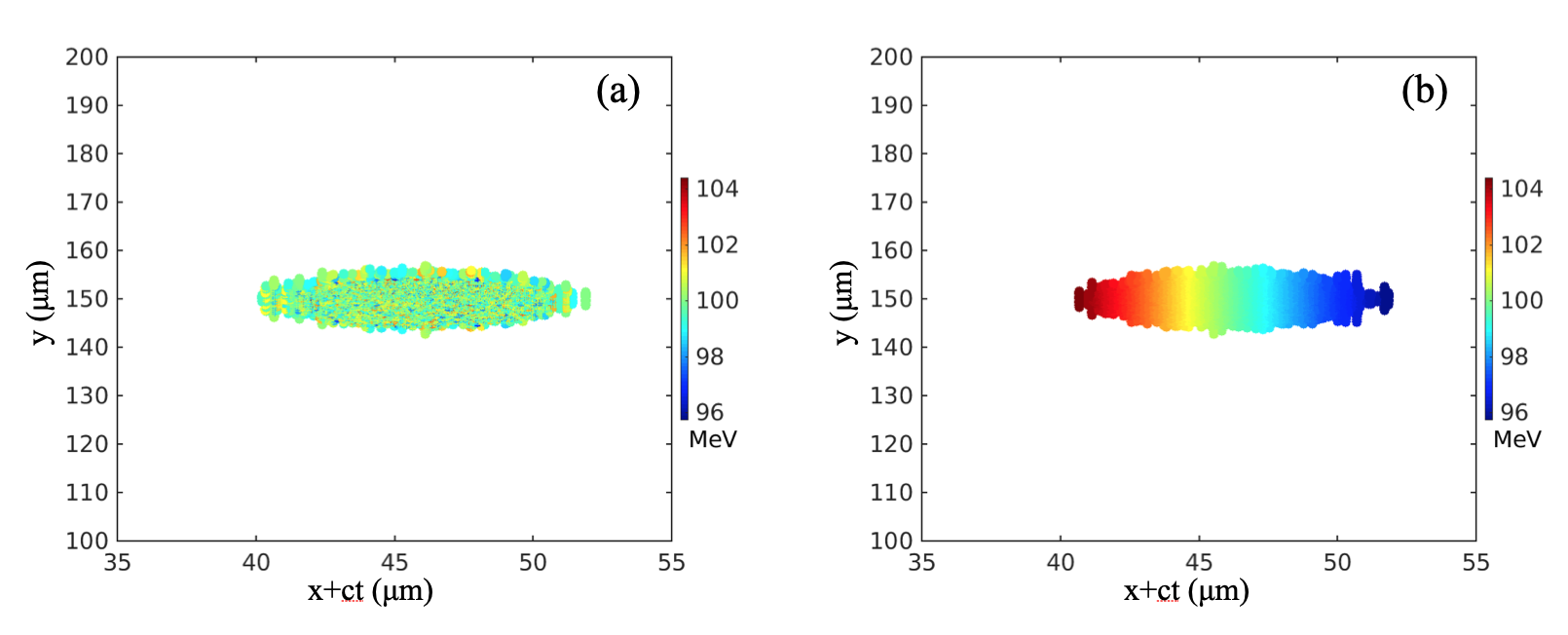}
\caption{\small a) Initial spatial distribution of an un-chirped electron beam and (b) a negatively chirped electron beam. The electron beam is propagating from right to left side.}\label{fig:3}
\end{figure}


\begin{figure}
\centering

\begin{subfigure}[t]{.45\textwidth}
\centering
\includegraphics[width=\linewidth]{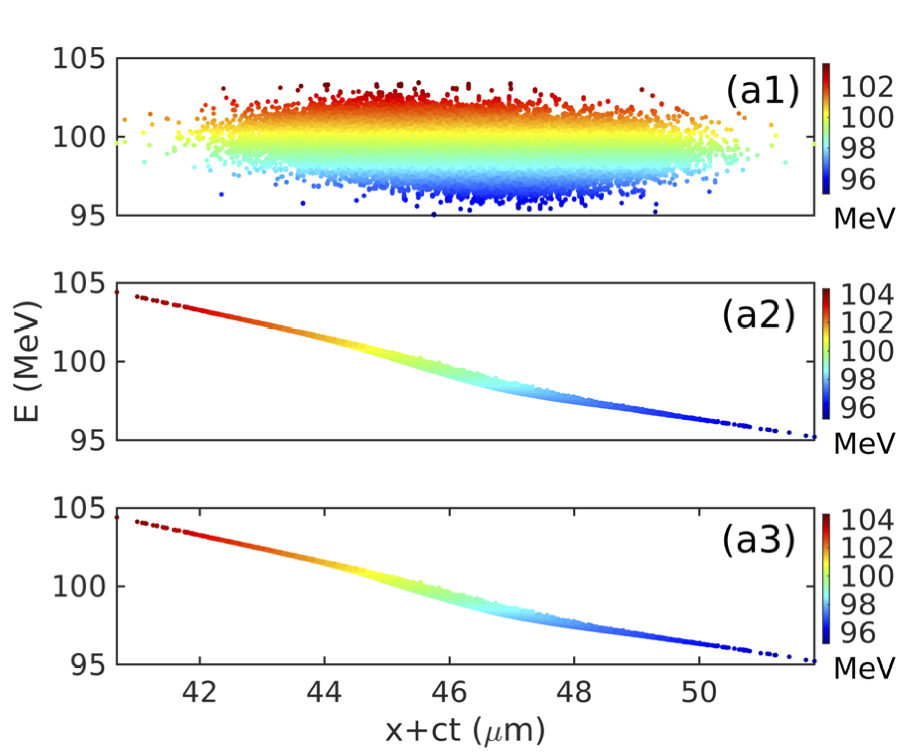}
\end{subfigure}
\hskip 20pt
\begin{subfigure}[t]{.45\textwidth}
\centering
\includegraphics[width=\linewidth]{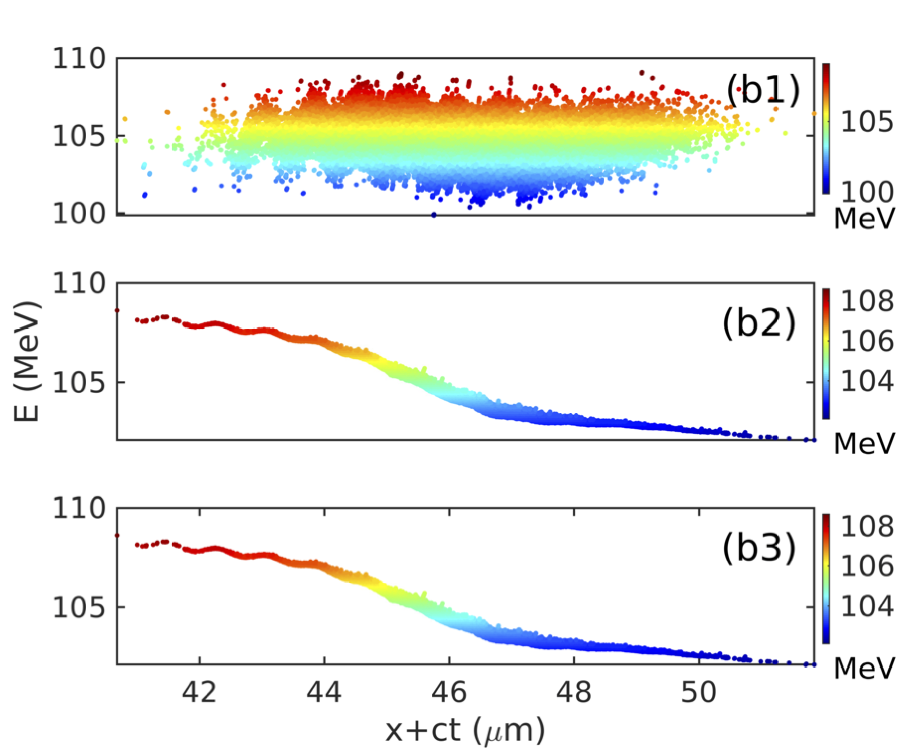}
\end{subfigure}


\begin{subfigure}[t]{.45\textwidth}
\centering
\vspace{-3pt}
\includegraphics[width=\linewidth]{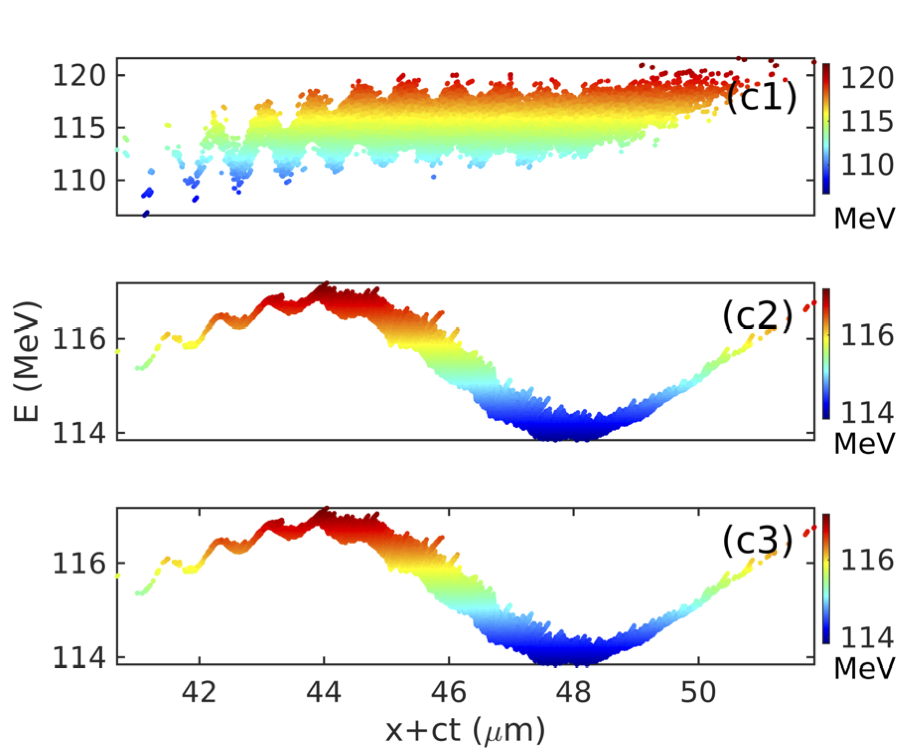}
\end{subfigure}
\hskip 20pt
\begin{subfigure}[t]{.45\textwidth}
\centering
\vspace{-3pt}
\includegraphics[width=\linewidth]{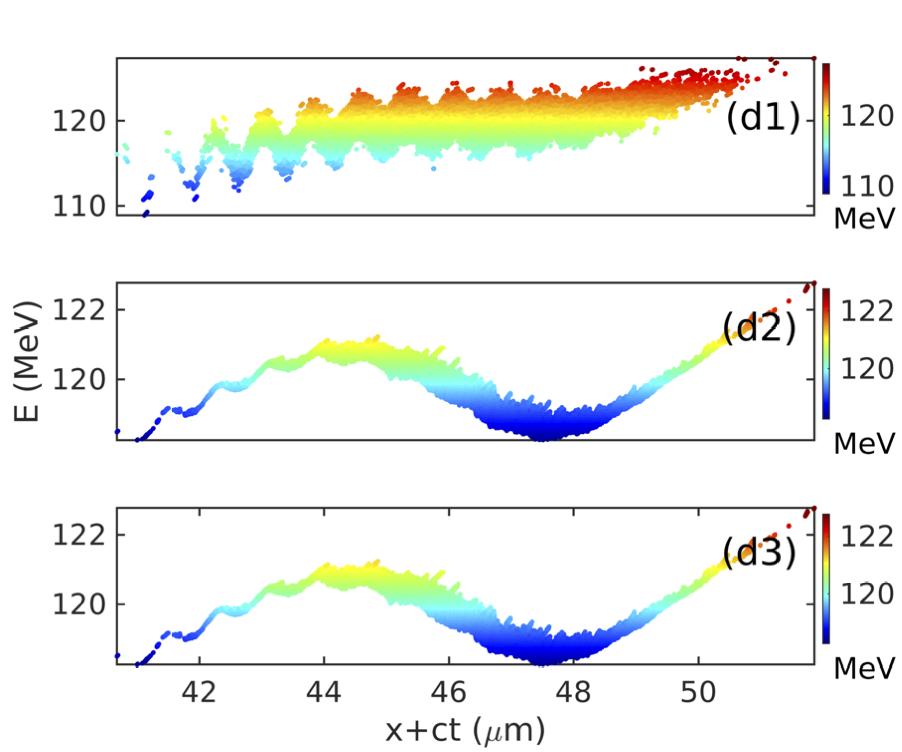}
\end{subfigure}
%


\begin{subfigure}[t]{.45\textwidth}
\centering
\vspace{10pt}
\includegraphics[width=\linewidth]{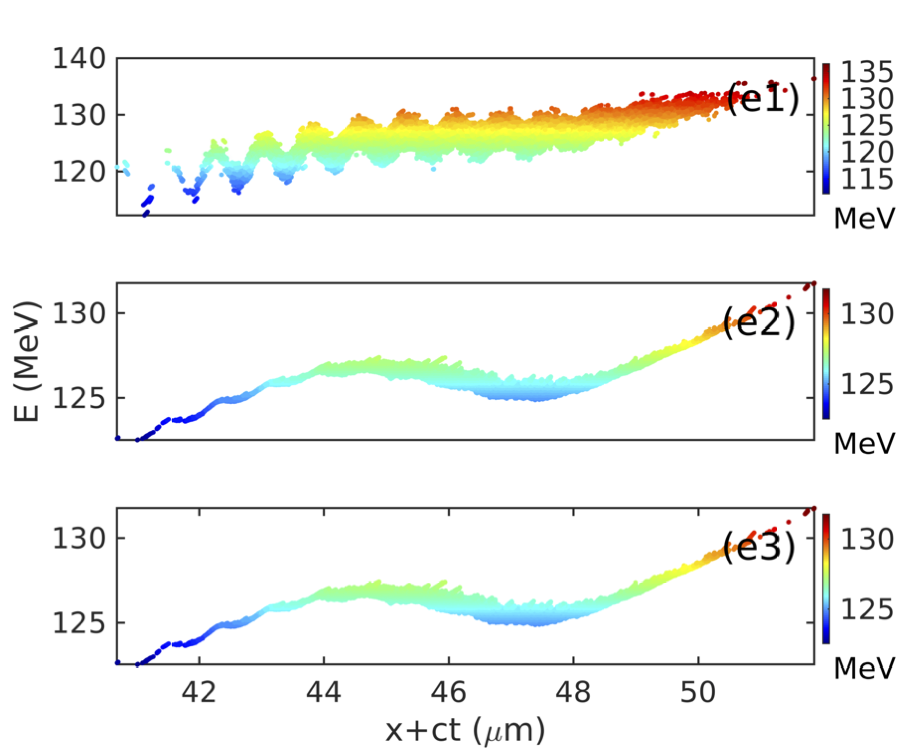}
\end{subfigure}
\hskip 20pt
\begin{subfigure}[t]{.45\textwidth}
\centering
\vspace{10pt}
\includegraphics[width=\linewidth]{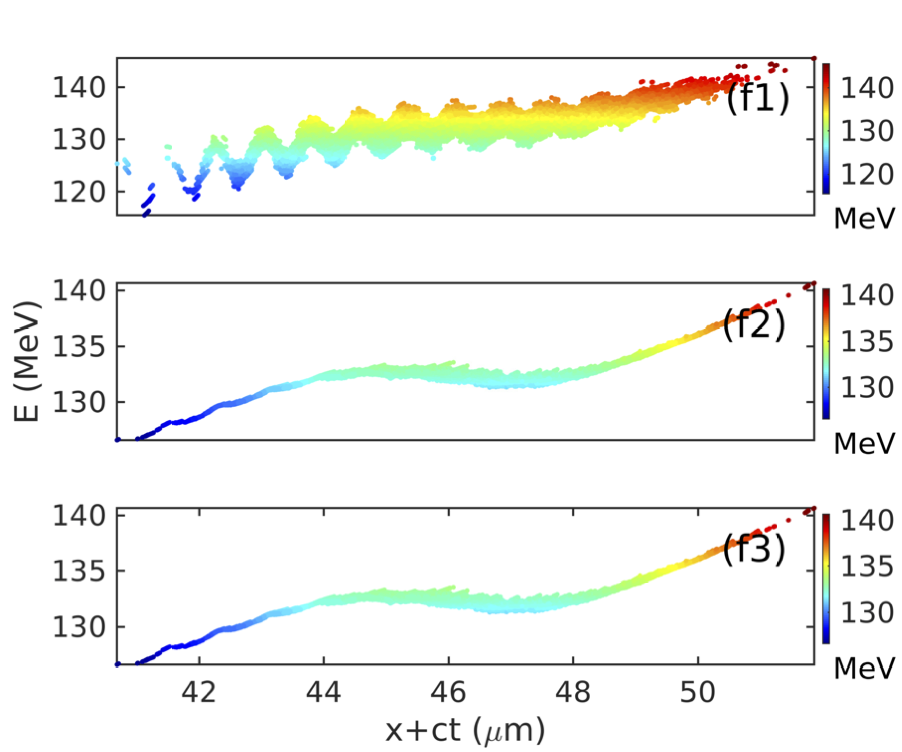}
\end{subfigure}
%
\begin{minipage}[t]{.8\textwidth}
\vspace{10pt}
\caption{\small Dynamic evolution of the phase-space of an initially unchirped electron beam (a1-f1), negative energy chirped electron beam (a2-f2) and negative energy chirped electron beam in longitudinally tapered plasma density (a3-f3).}
\end{minipage}

\end{figure}


\begin{figure}
\centering

\begin{subfigure}[t]{.45\textwidth}
\centering
\includegraphics[width=\linewidth]{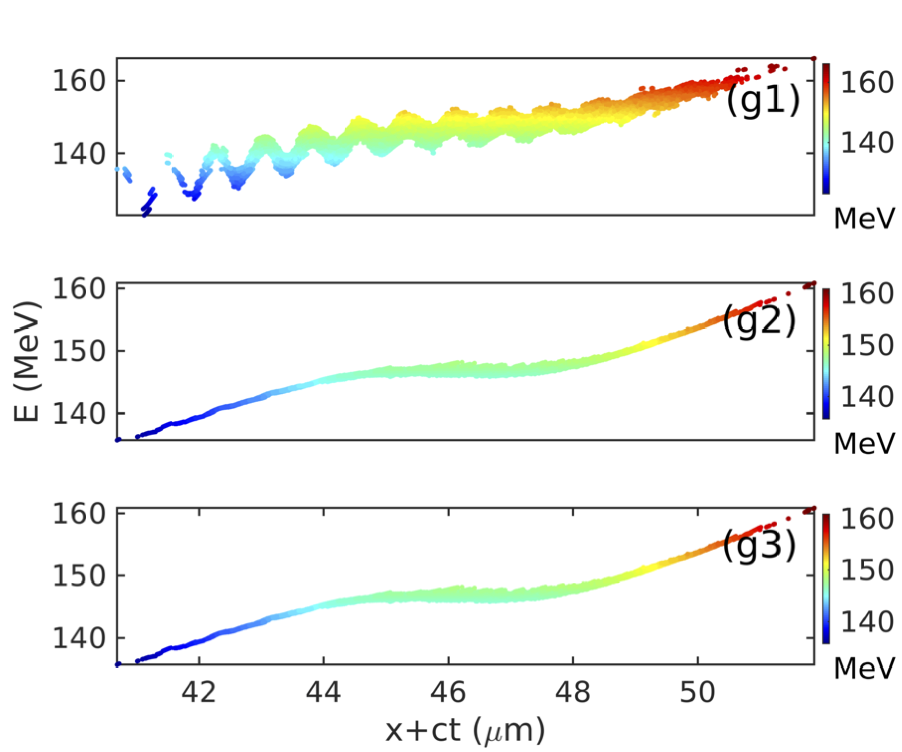}
\end{subfigure}
\hskip 20pt
\begin{subfigure}[t]{.45\textwidth}
\centering
\includegraphics[width=\linewidth]{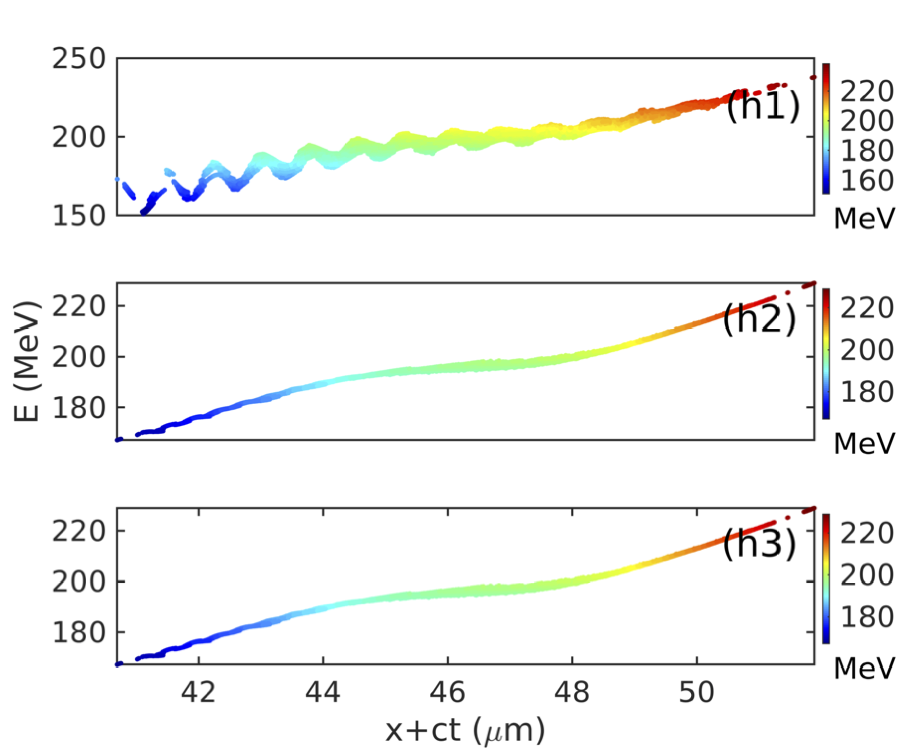}
\end{subfigure}


\begin{subfigure}[t]{.45\textwidth}
\centering
\vspace{-3pt}
\includegraphics[width=\linewidth]{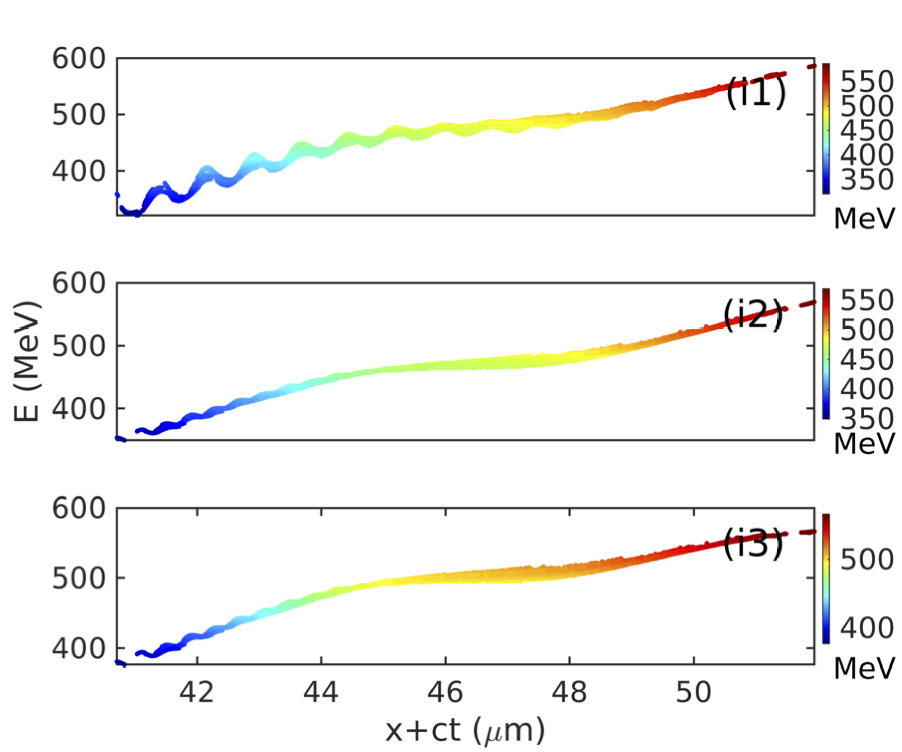}
\end{subfigure}
\hskip 20pt
\begin{subfigure}[t]{.45\textwidth}
\centering
\vspace{-3pt}
\includegraphics[width=\linewidth]{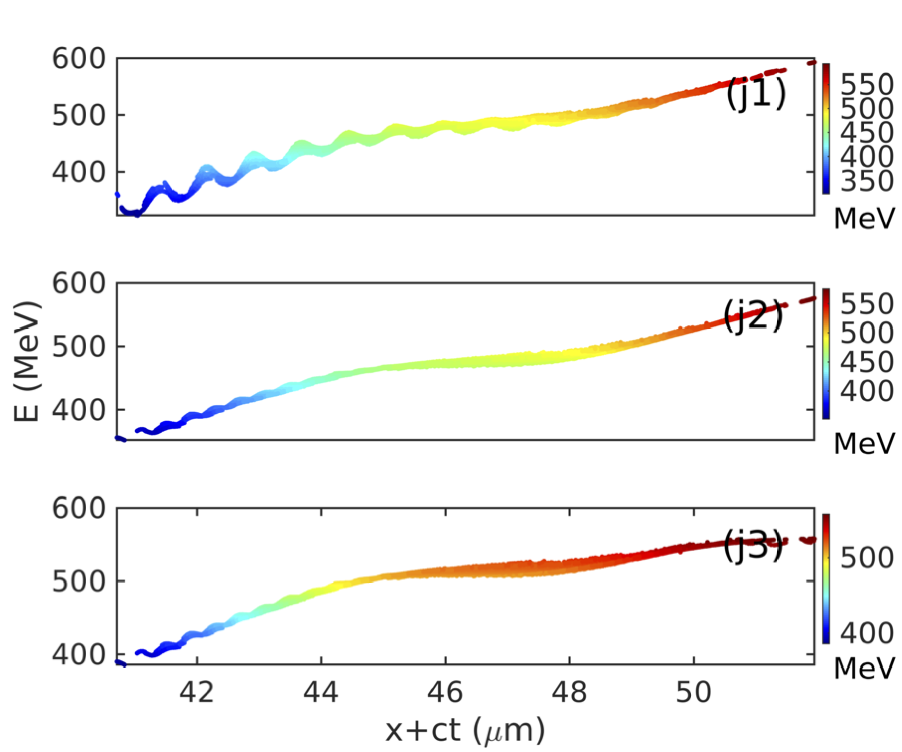}
\end{subfigure}
%


\begin{subfigure}[t]{.45\textwidth}
\centering
\vspace{10pt}
\includegraphics[width=\linewidth]{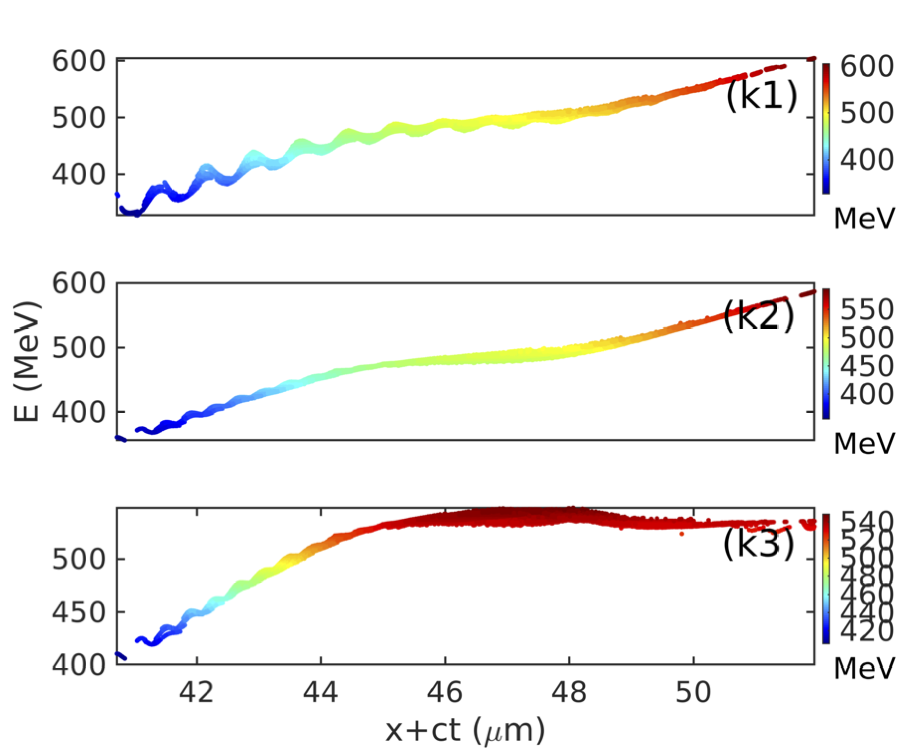}
\end{subfigure}
\hskip 20pt
\begin{subfigure}[t]{.45\textwidth}
\centering
\vspace{10pt}
\includegraphics[width=\linewidth]{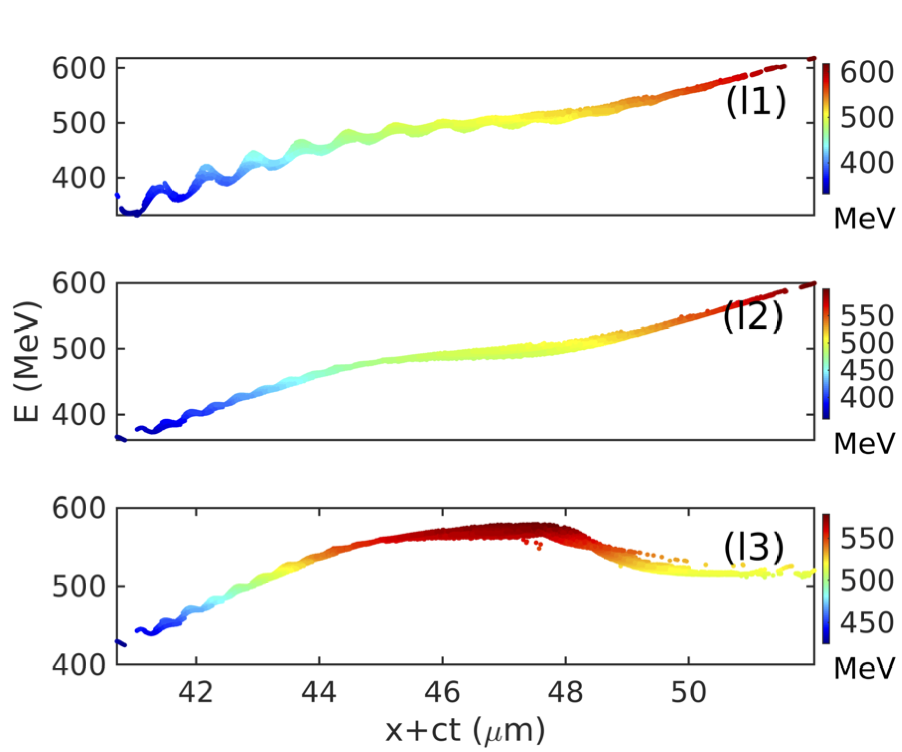}
\end{subfigure}
%
\begin{minipage}[t]{.8\textwidth}
\vspace{10pt}
\caption{\small Dynamic evolution of the phase-space of an initially unchirped electron beam (g1-l1), negative energy chirped electron beam (g2-l2) and negative energy chirped electron beam in longitudinally tapered plasma density (g3-l3).}
\end{minipage}

\end{figure}


\begin{figure}
\centering

\begin{subfigure}[t]{.45\textwidth}
\centering
\includegraphics[width=\linewidth]{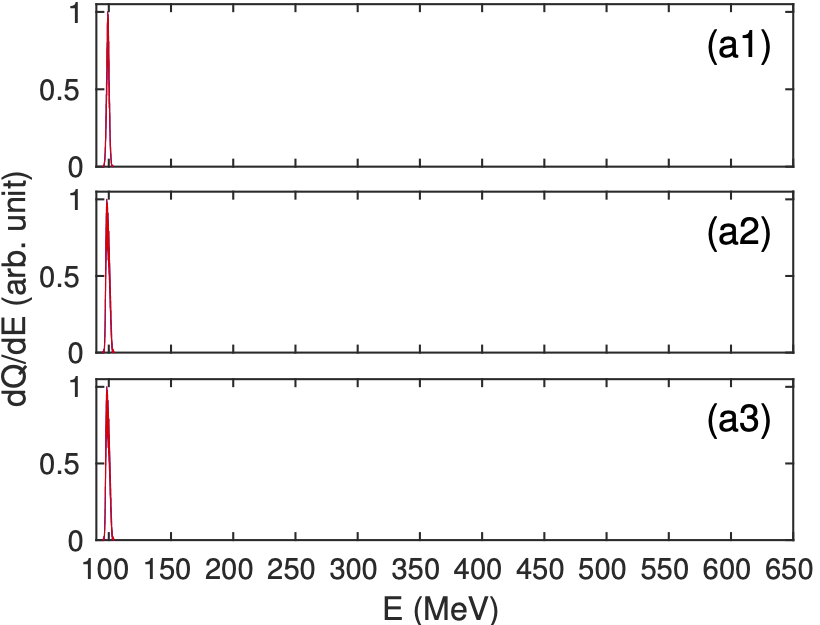}
\end{subfigure}
\hskip 20pt
\begin{subfigure}[t]{.45\textwidth}
\centering
\includegraphics[width=\linewidth]{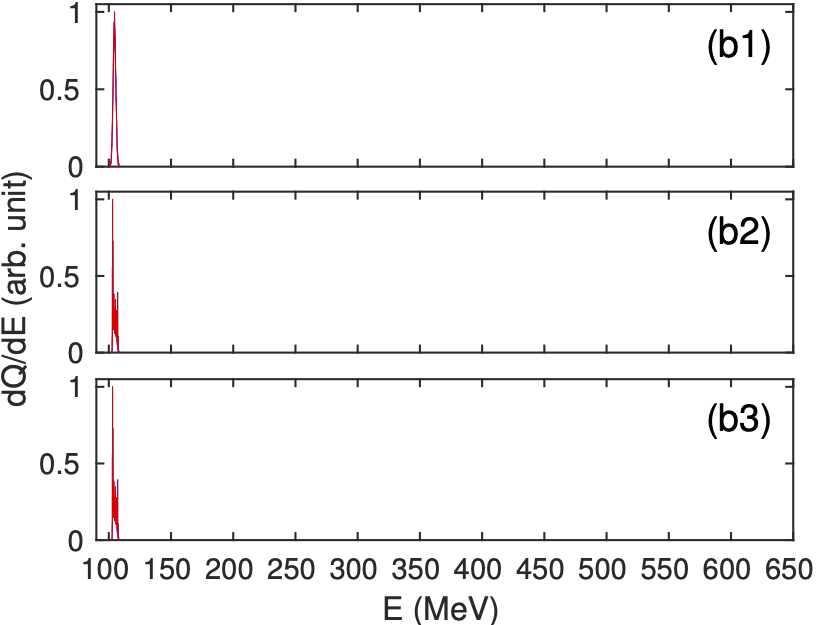}
\end{subfigure}


\begin{subfigure}[t]{.45\textwidth}
\centering
\vspace{5pt}
\includegraphics[width=\linewidth]{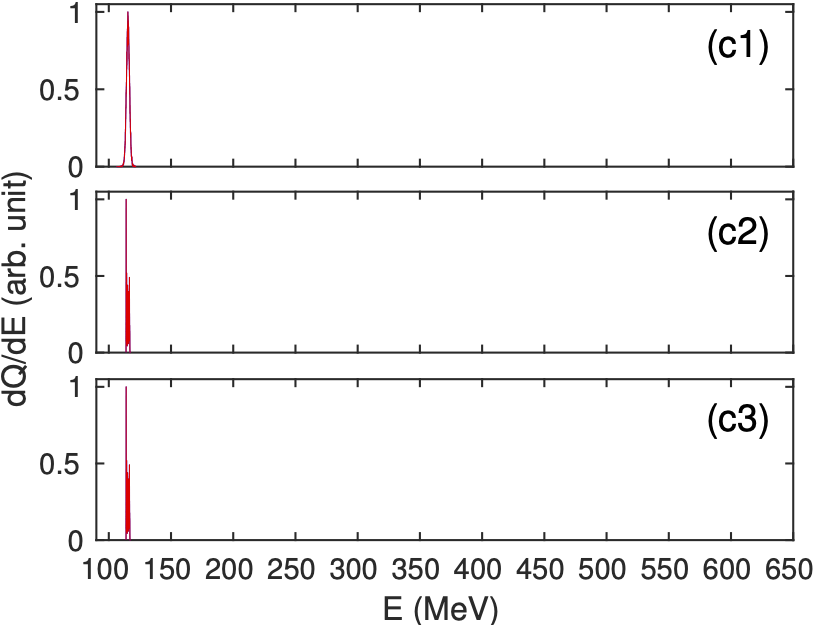}
\end{subfigure}
\hskip 20pt
\begin{subfigure}[t]{.45\textwidth}
\centering
\vspace{5pt}
\includegraphics[width=\linewidth]{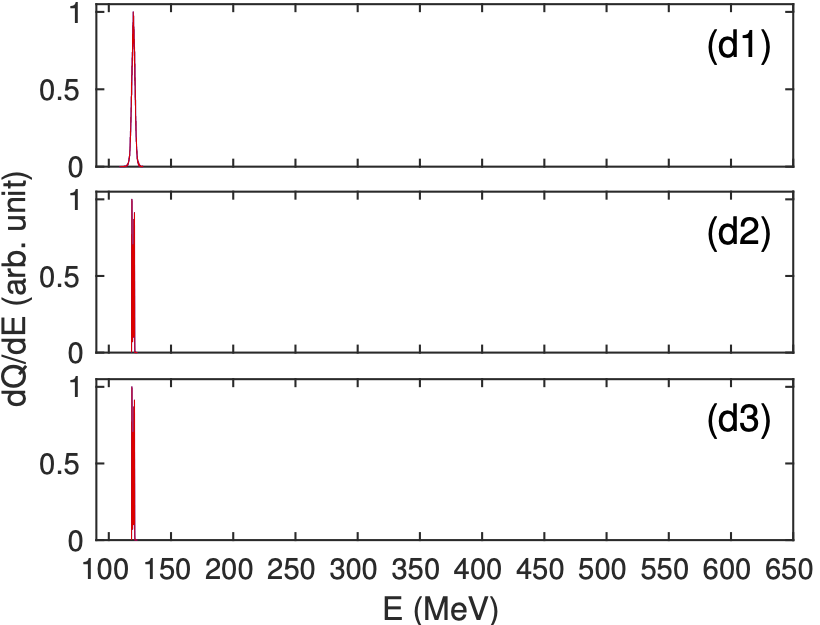}
\end{subfigure}
%


\begin{subfigure}[t]{.45\textwidth}
\centering
\vspace{25pt}
\includegraphics[width=\linewidth]{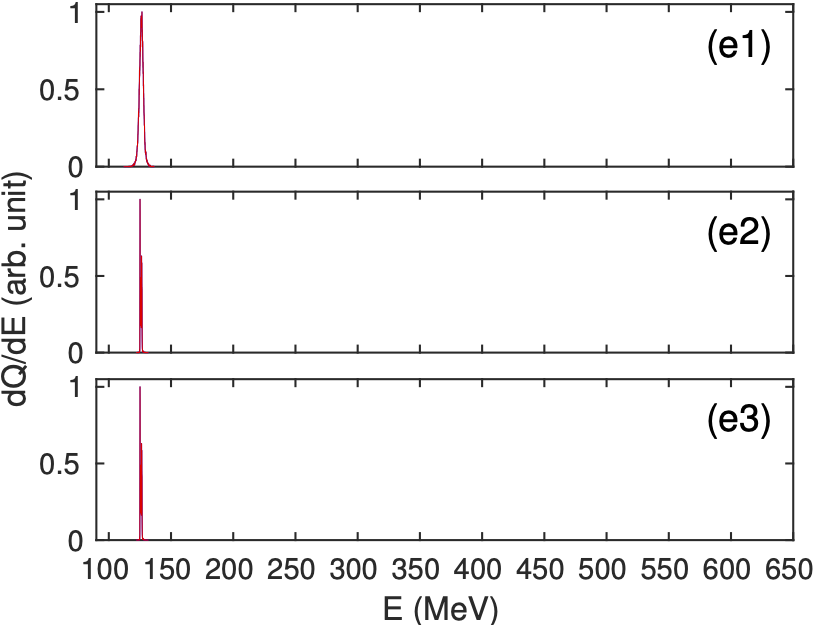}
\end{subfigure}
\hskip 20pt
\begin{subfigure}[t]{.45\textwidth}
\centering
\vspace{25pt}
\includegraphics[width=\linewidth]{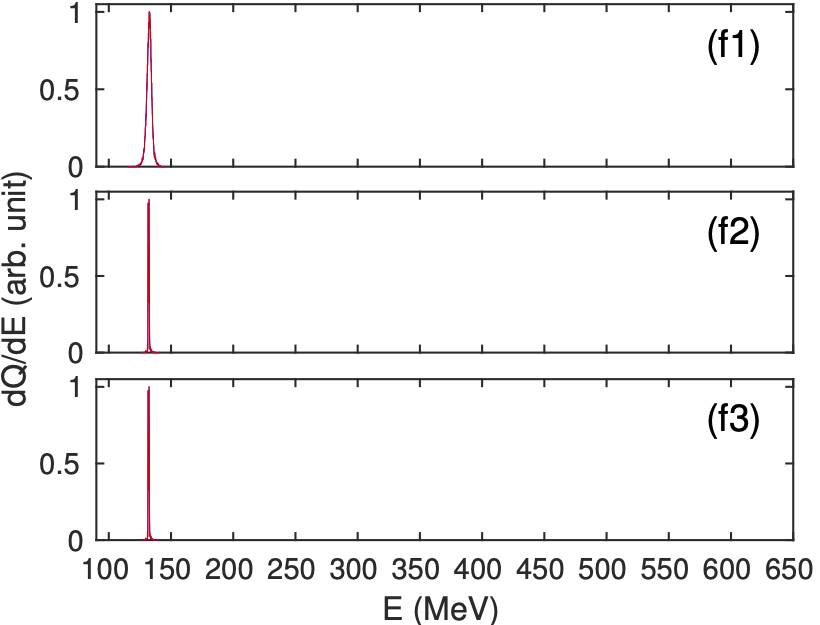}
\end{subfigure}
%
\begin{minipage}[t]{.8\textwidth}
\vspace{20pt}
\caption{\small Dynamic evolution of the energy spectrum of an initially unchirped electron beam (a1-f1), negative energy chirped electron beam (a2-f2) and negative energy chirped electron beam in longitudinally tapered plasma density (a3-f3).}
\end{minipage}

\end{figure}


\begin{figure}
\centering

\begin{subfigure}[t]{.45\textwidth}
\centering
\includegraphics[width=\linewidth]{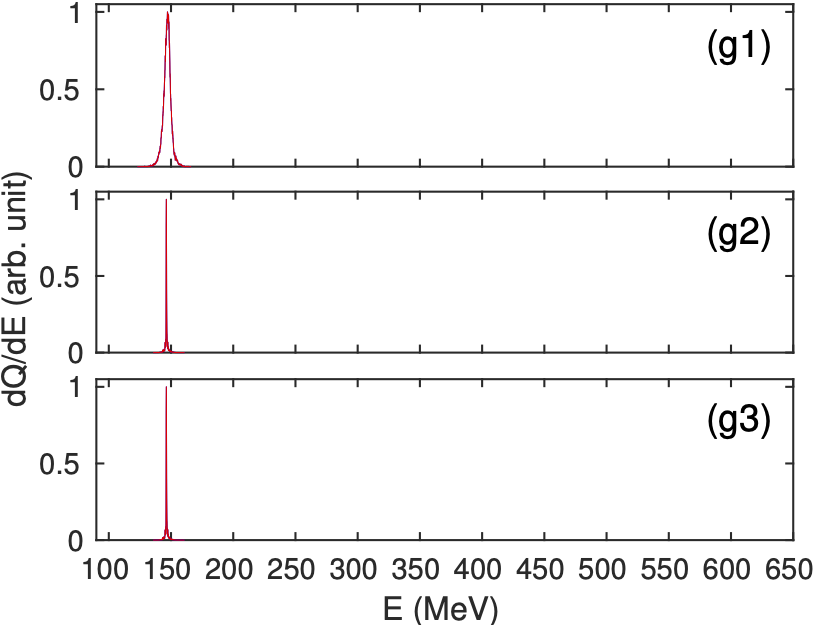}
\end{subfigure}
\hskip 20pt
\begin{subfigure}[t]{.45\textwidth}
\centering
\includegraphics[width=\linewidth]{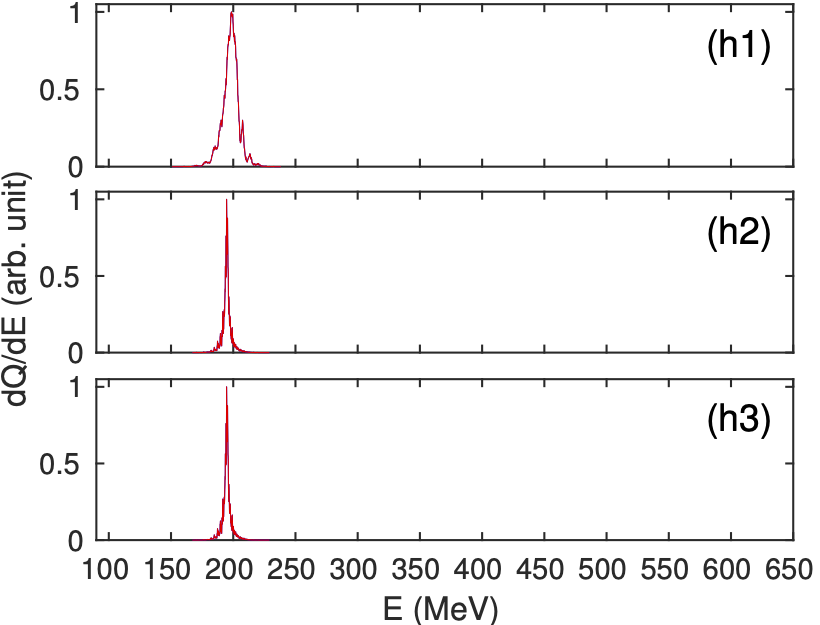}
\end{subfigure}


\begin{subfigure}[t]{.45\textwidth}
\centering
\vspace{5pt}
\includegraphics[width=\linewidth]{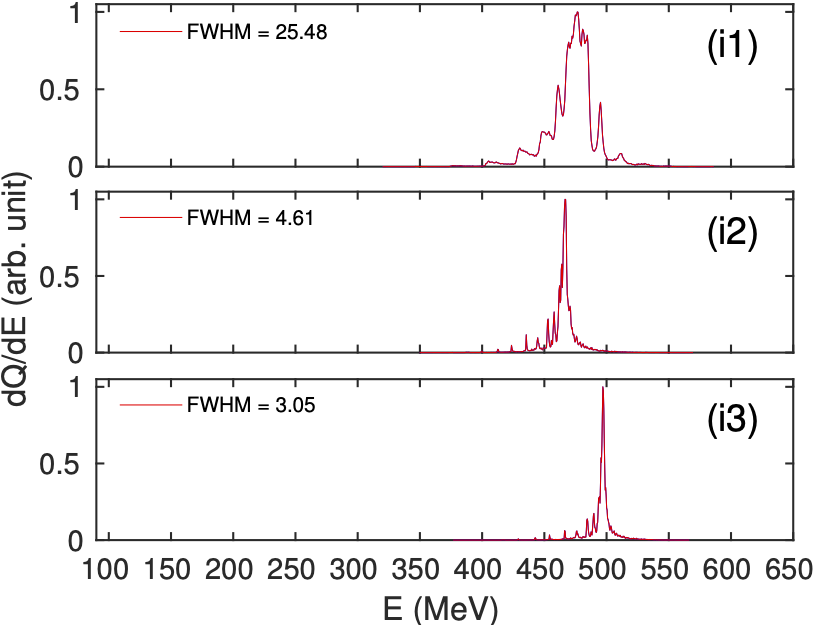}
\end{subfigure}
\hskip 20pt
\begin{subfigure}[t]{.45\textwidth}
\centering
\vspace{5pt}
\includegraphics[width=\linewidth]{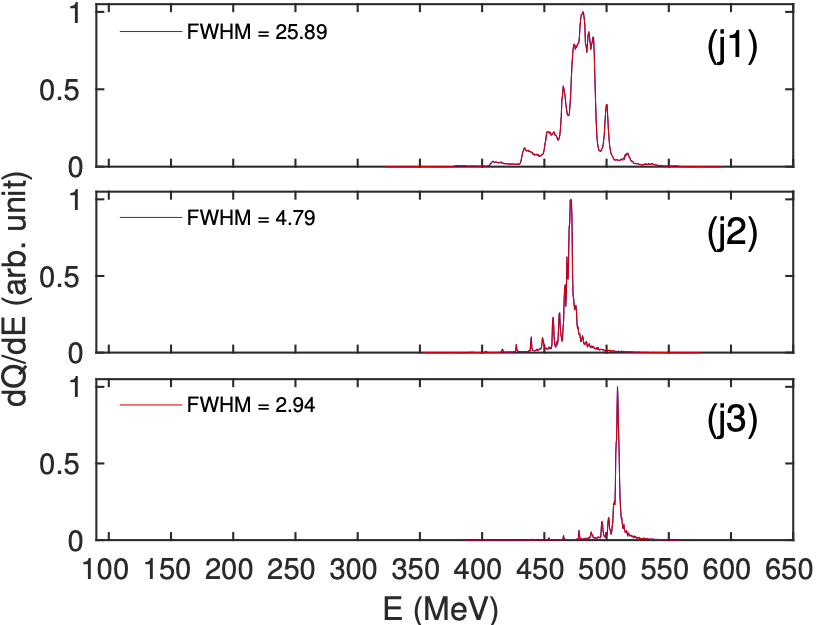}
\end{subfigure}
%


\begin{subfigure}[t]{.45\textwidth}
\centering
\vspace{25pt}
\includegraphics[width=\linewidth]{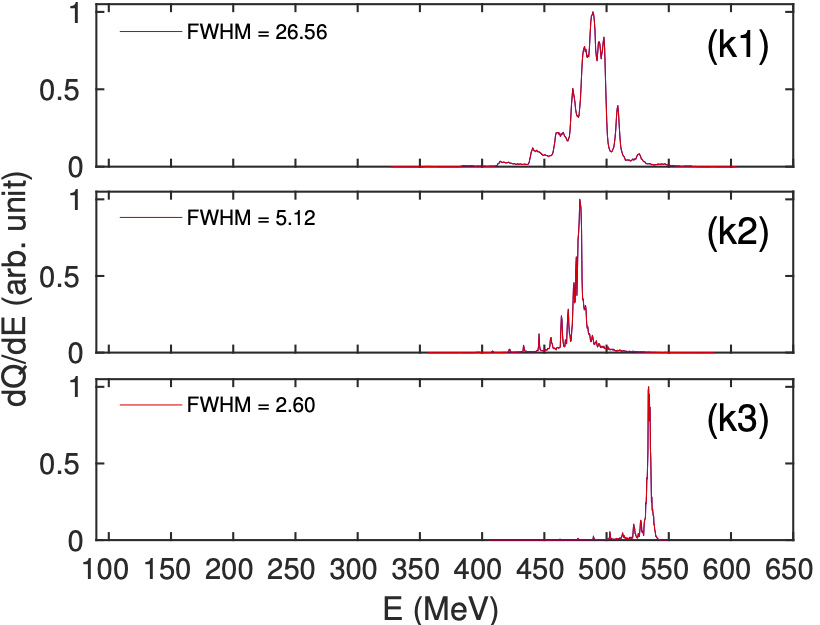}
\end{subfigure}
\hskip 20pt
\begin{subfigure}[t]{.45\textwidth}
\centering
\vspace{25pt}
\includegraphics[width=\linewidth]{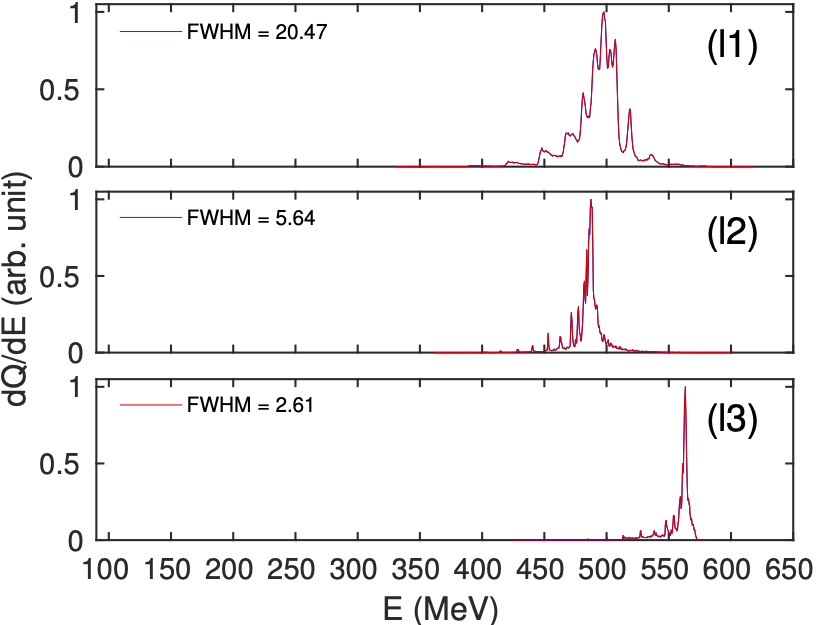}
\end{subfigure}
%
\begin{minipage}[t]{.8\textwidth}
\vspace{20pt}
\caption{\small Dynamic evolution of the energy spectrum of an initially unchirped electron beam (g1-l1), negative energy chirped electron beam (g2-l2) and negative energy chirped electron beam in longitudinally tapered plasma density (g3-l3).}
\end{minipage}

\end{figure}


\begin{figure}
\centering
\includegraphics[width=0.6\linewidth]{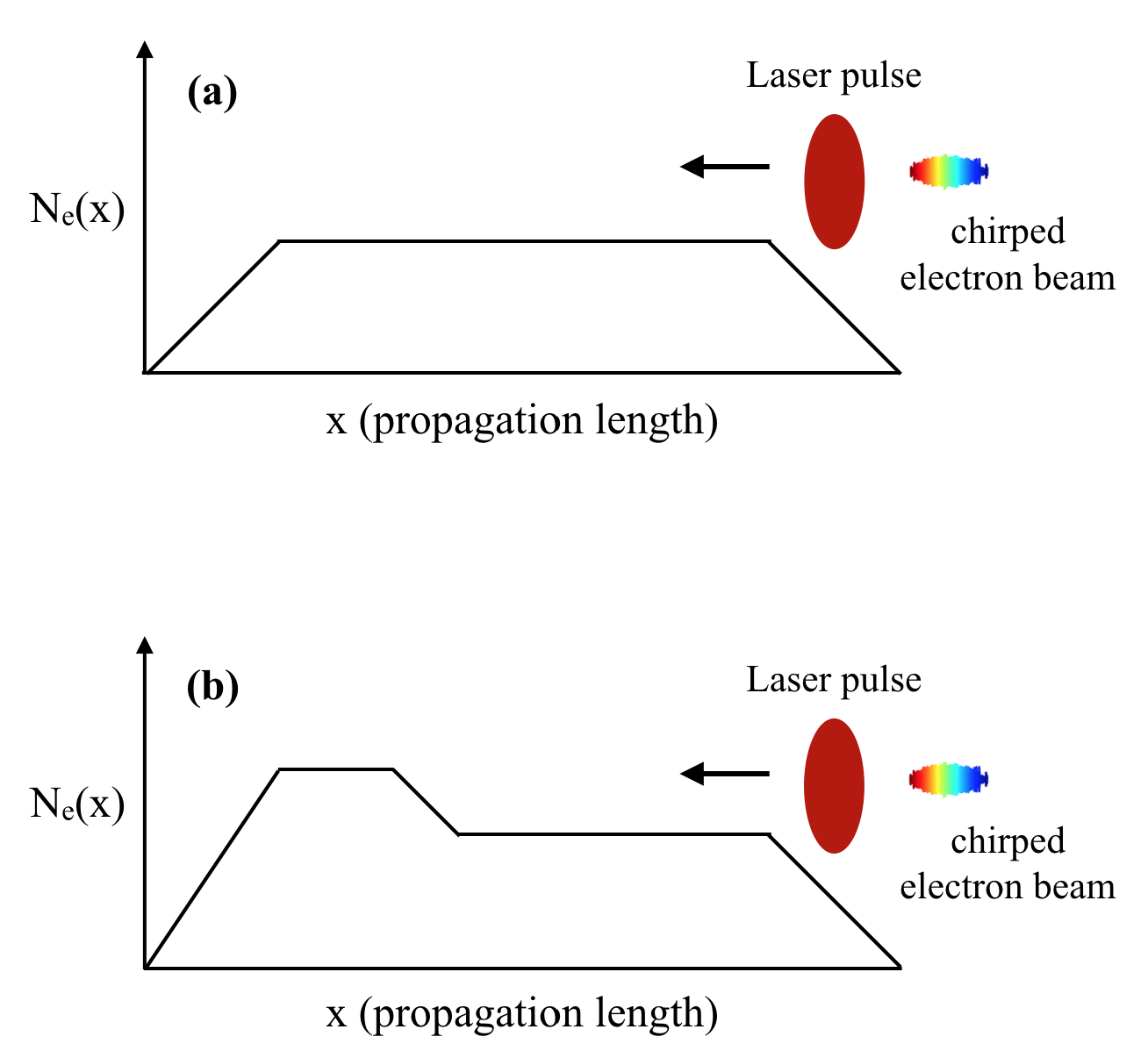}
\caption{\small 1D profile of longitudinal plasma density. (a) Uniform (except rising and falling edges at the entrance and exist of the plasma medium) and (b) longitudinal tapering of the plasma density in order to further narrowing down of the energy spectrum.} 
\end{figure}


\begin{figure}
\centering

\begin{subfigure}[t]{.45\textwidth}
\centering
\includegraphics[width=\linewidth]{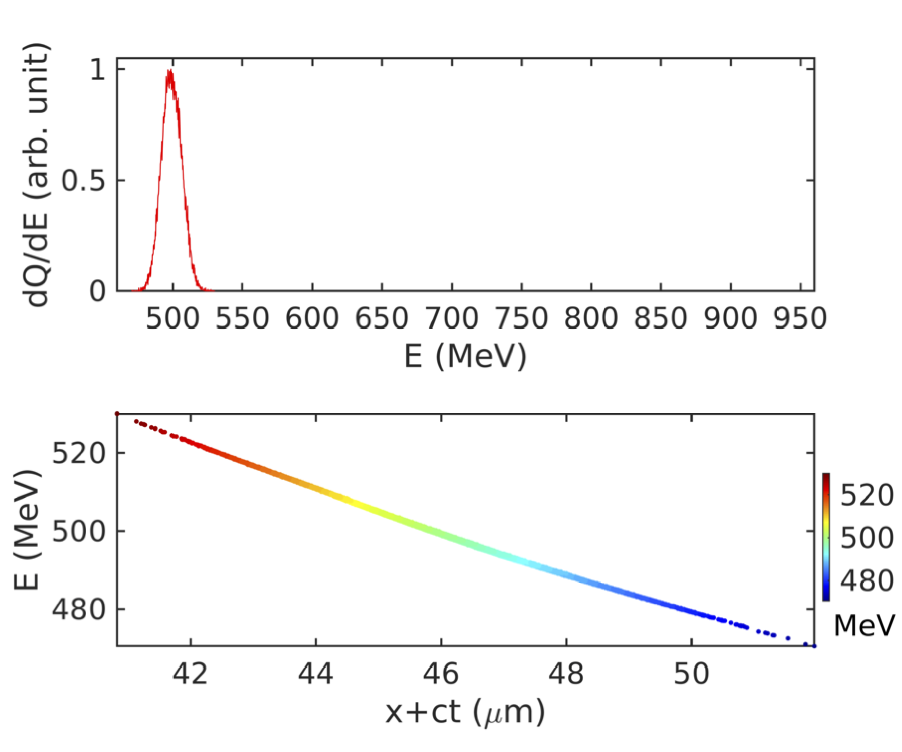}
\caption{}\label{fig:fig_a}
\end{subfigure}
\hskip 20pt
\begin{subfigure}[t]{.45\textwidth}
\centering
\includegraphics[width=\linewidth]{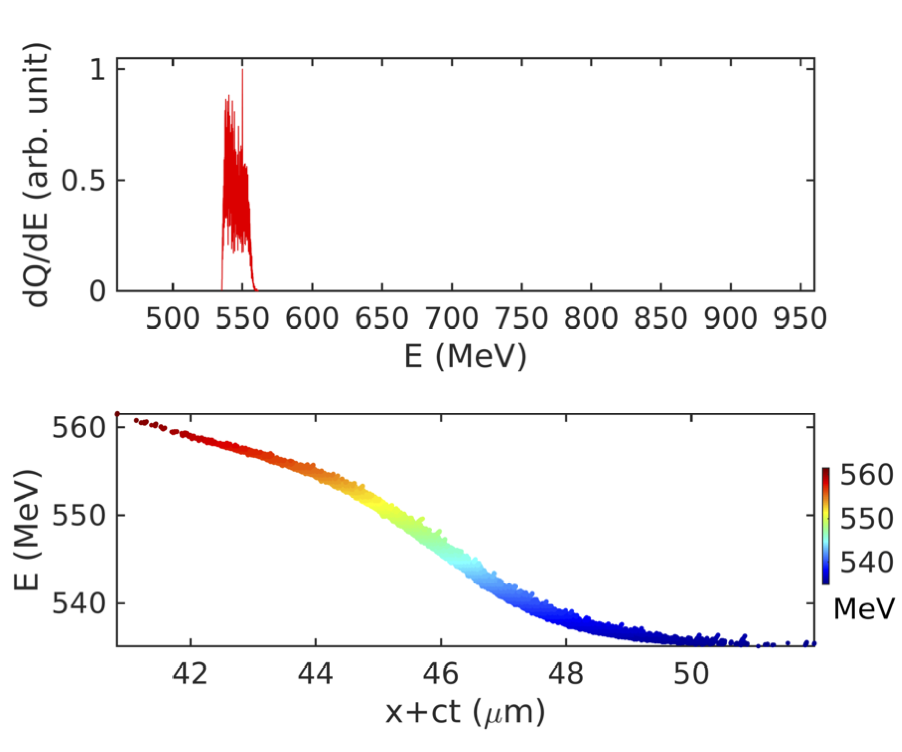}
\caption{}\label{fig:fig_b}
\end{subfigure}


\begin{subfigure}[t]{.45\textwidth}
\centering
\vspace{5pt}
\includegraphics[width=\linewidth]{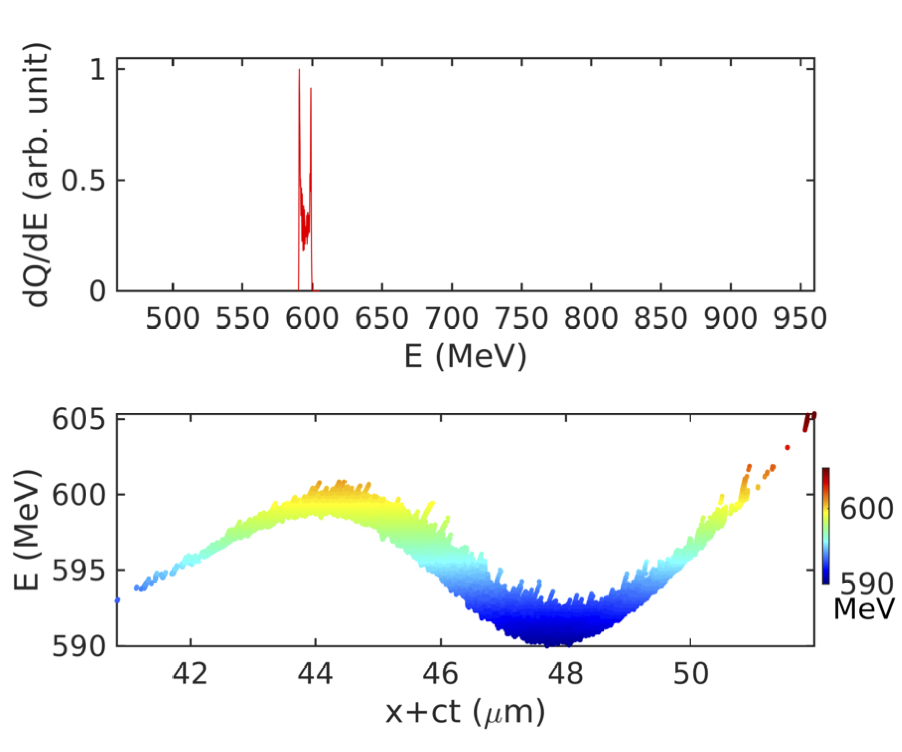}
\caption{}\label{fig:fig_c}
\end{subfigure}
\hskip 20pt
\begin{subfigure}[t]{.45\textwidth}
\centering
\vspace{5pt}
\includegraphics[width=\linewidth]{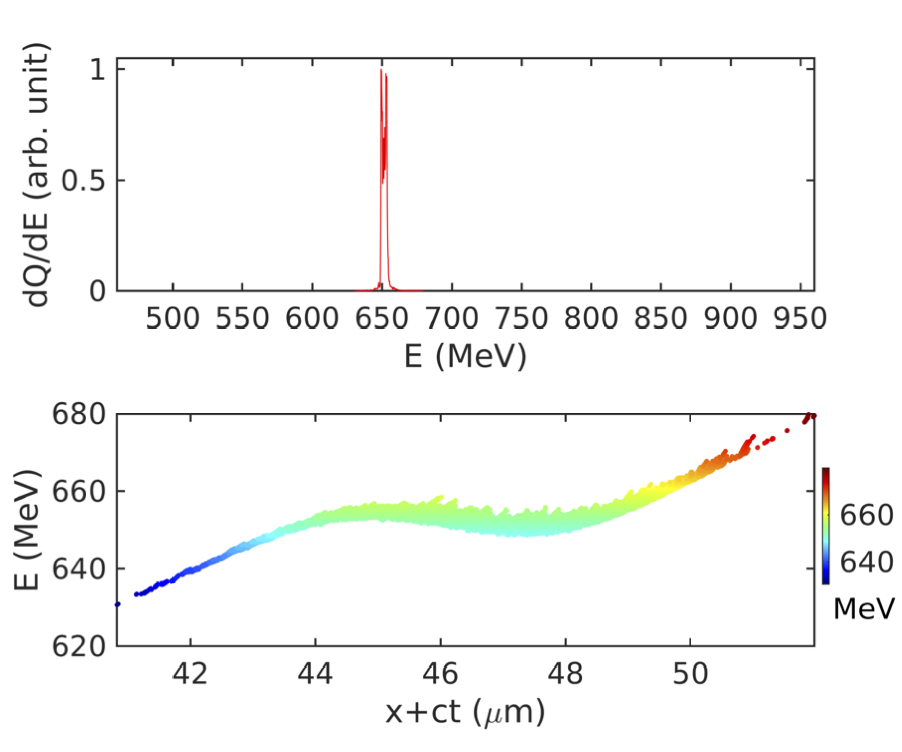}
\caption{}\label{fig:fig_d}
\end{subfigure}
%


\begin{subfigure}[t]{.45\textwidth}
\centering
\vspace{25pt}
\includegraphics[width=\linewidth]{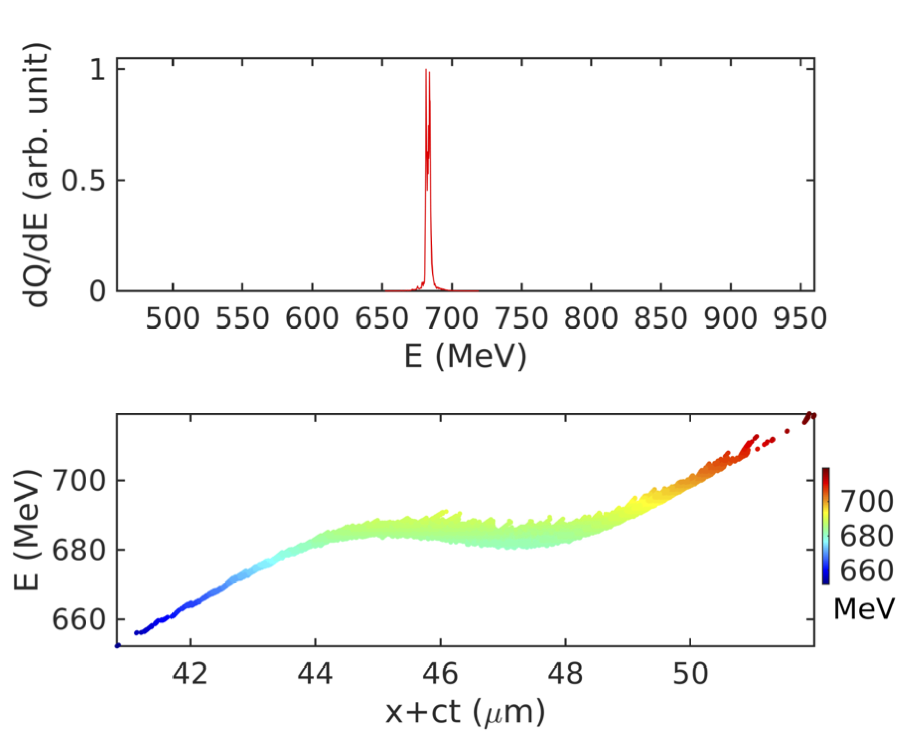}
\caption{}\label{fig:fig_e}
\end{subfigure}
\hskip 20pt
\begin{subfigure}[t]{.45\textwidth}
\centering
\vspace{25pt}
\includegraphics[width=\linewidth]{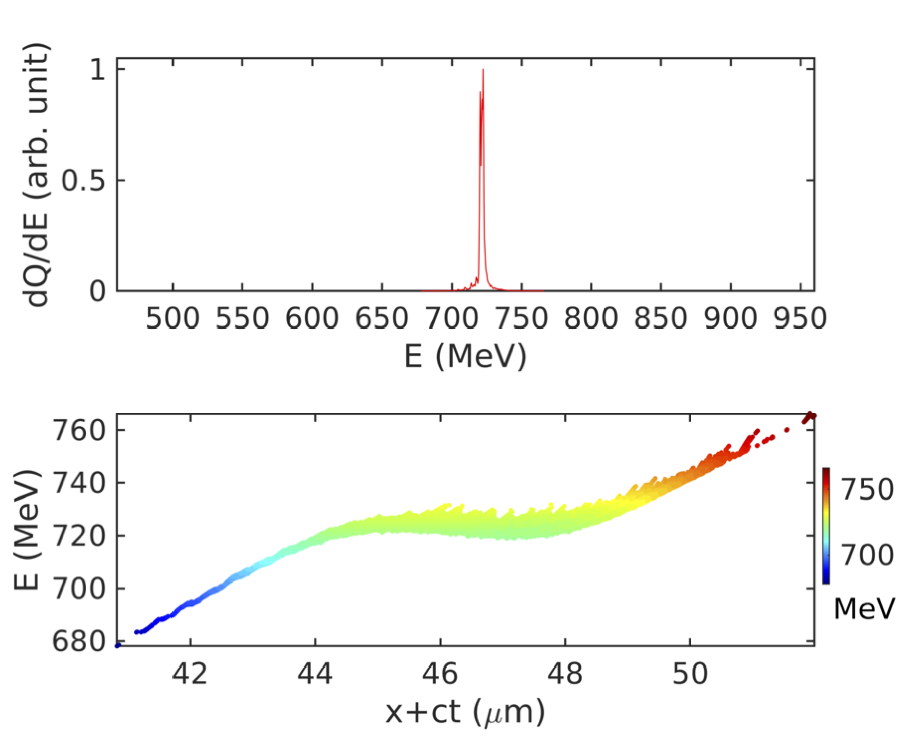}
\caption{}\label{fig:fig_f}
\end{subfigure}
%
\begin{minipage}[t]{.8\textwidth}
\vspace{20pt}
\caption{\small Dynamic evolution of the energy spectrum and longitudinal phase-space of an initially negative energy chirped electron beam with an initial enegy of 500 MeV.}
\end{minipage}

\end{figure}


\begin{figure}
\centering

\begin{subfigure}[t]{.45\textwidth}
\centering
\includegraphics[width=\linewidth]{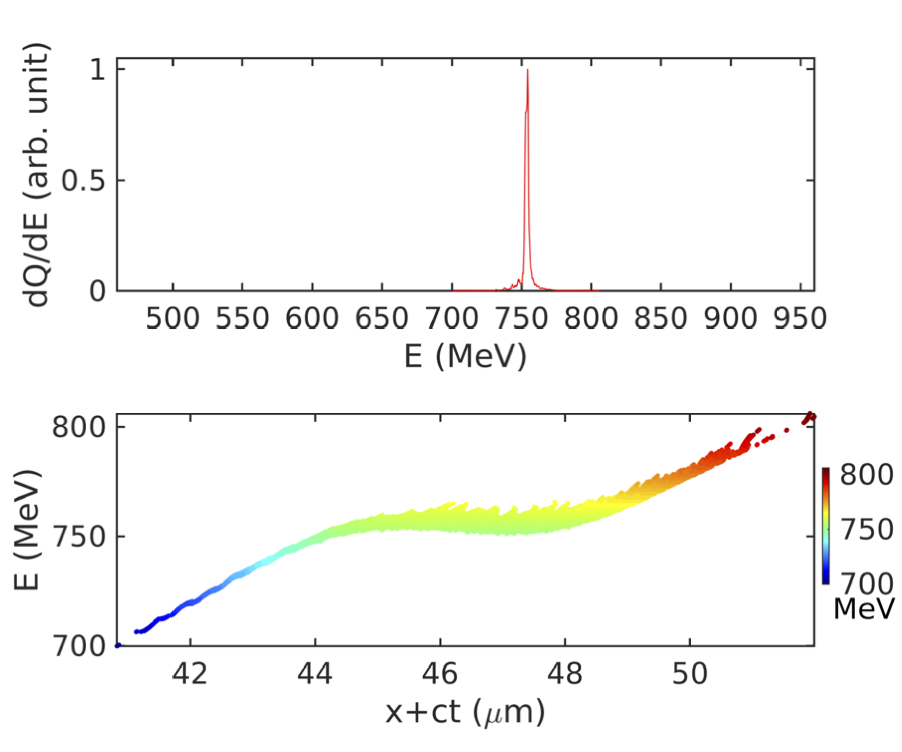}
\caption{}\label{fig:fig_a}
\end{subfigure}
\hskip 20pt
\begin{subfigure}[t]{.45\textwidth}
\centering
\includegraphics[width=\linewidth]{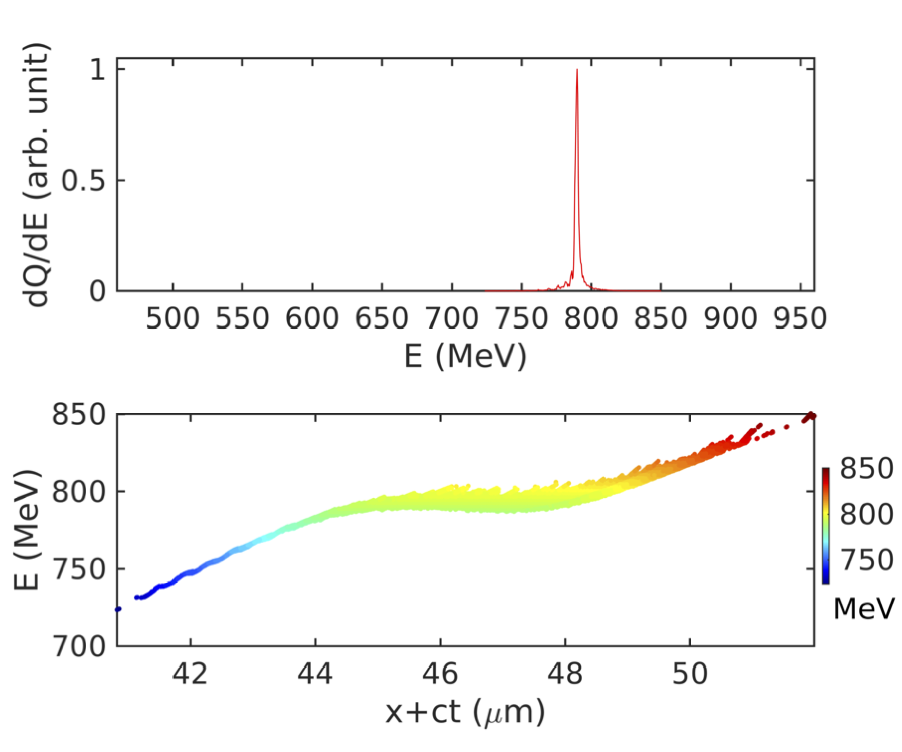}
\caption{}\label{fig:fig_b}
\end{subfigure}


\begin{subfigure}[t]{.45\textwidth}
\centering
\vspace{5pt}
\includegraphics[width=\linewidth]{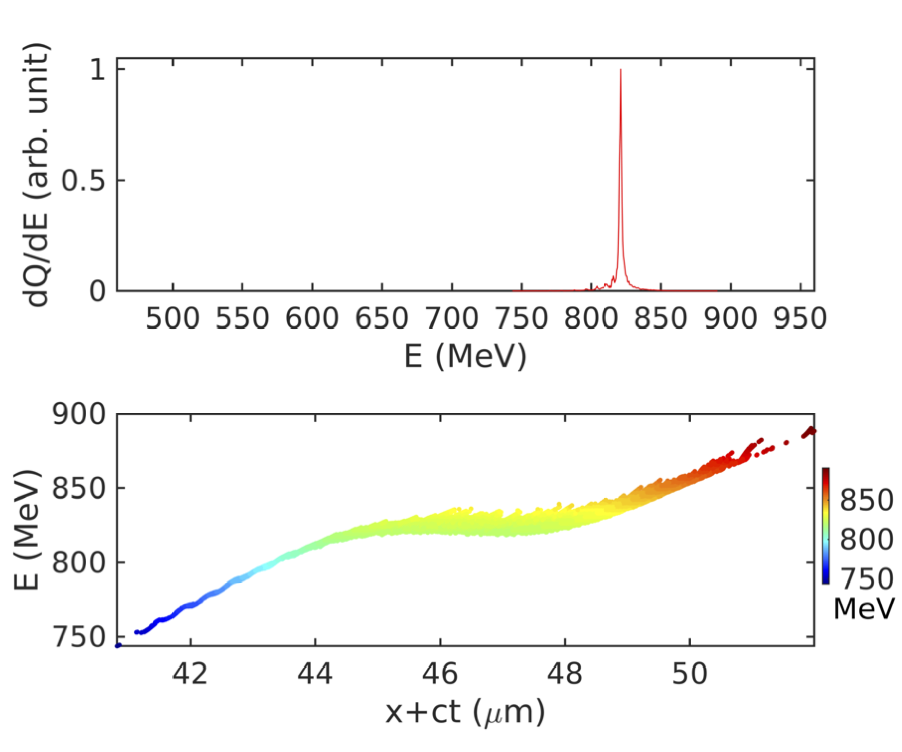}
\caption{}\label{fig:fig_c}
\end{subfigure}
\hskip 20pt
\begin{subfigure}[t]{.45\textwidth}
\centering
\vspace{5pt}
\includegraphics[width=\linewidth]{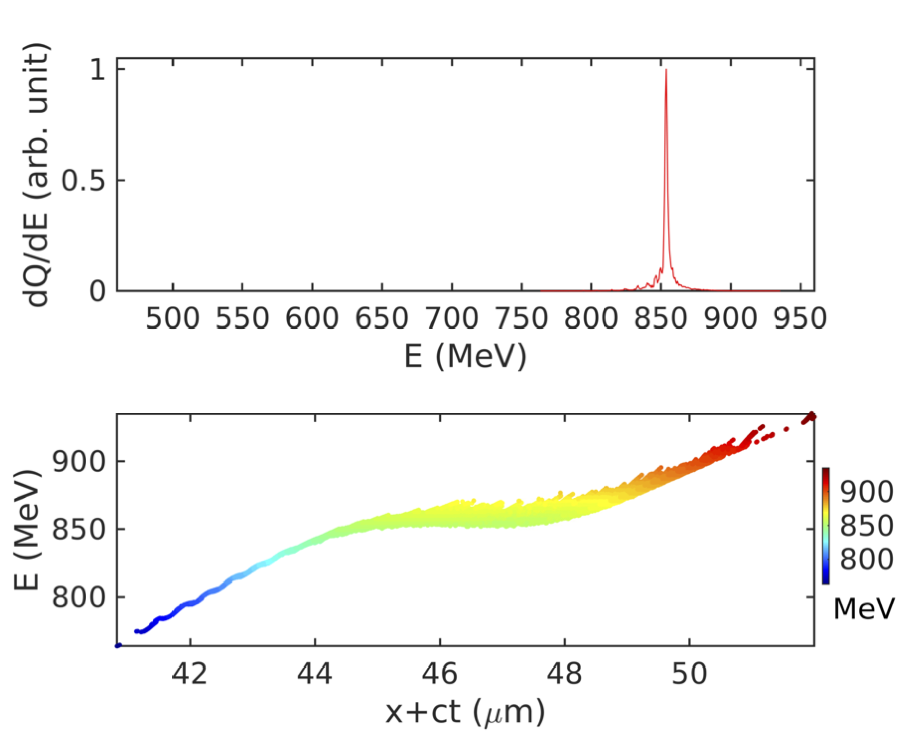}
\caption{}\label{fig:fig_d}
\end{subfigure}
%


\begin{subfigure}[t]{.45\textwidth}
\centering
\vspace{25pt}
\includegraphics[width=\linewidth]{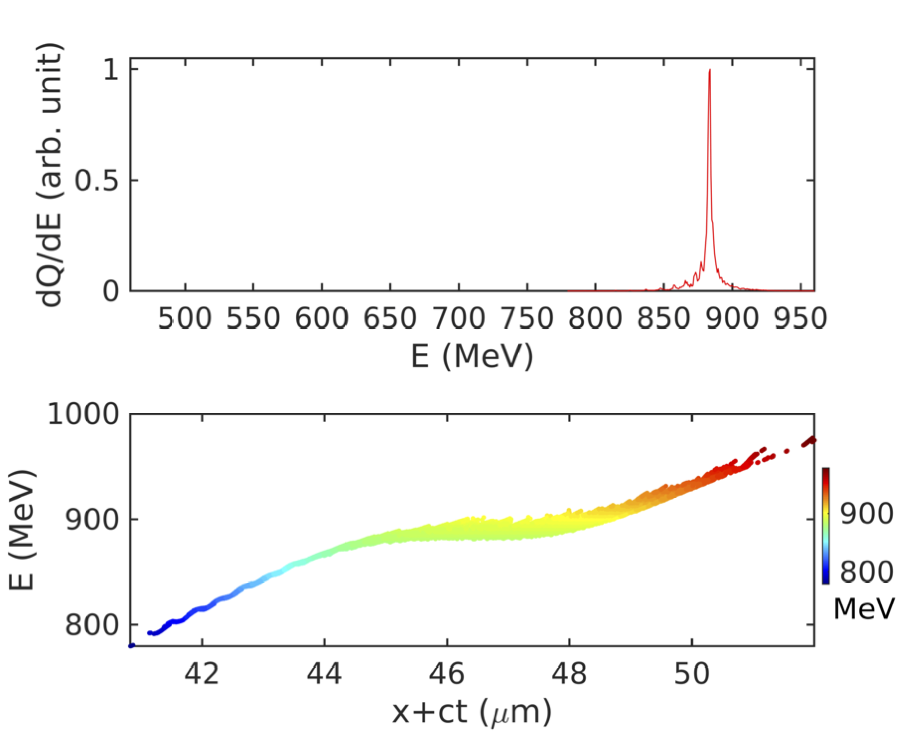}
\caption{}\label{fig:fig_e}
\end{subfigure}
\hskip 20pt
\begin{subfigure}[t]{.45\textwidth}
\centering
\vspace{25pt}
\includegraphics[width=\linewidth]{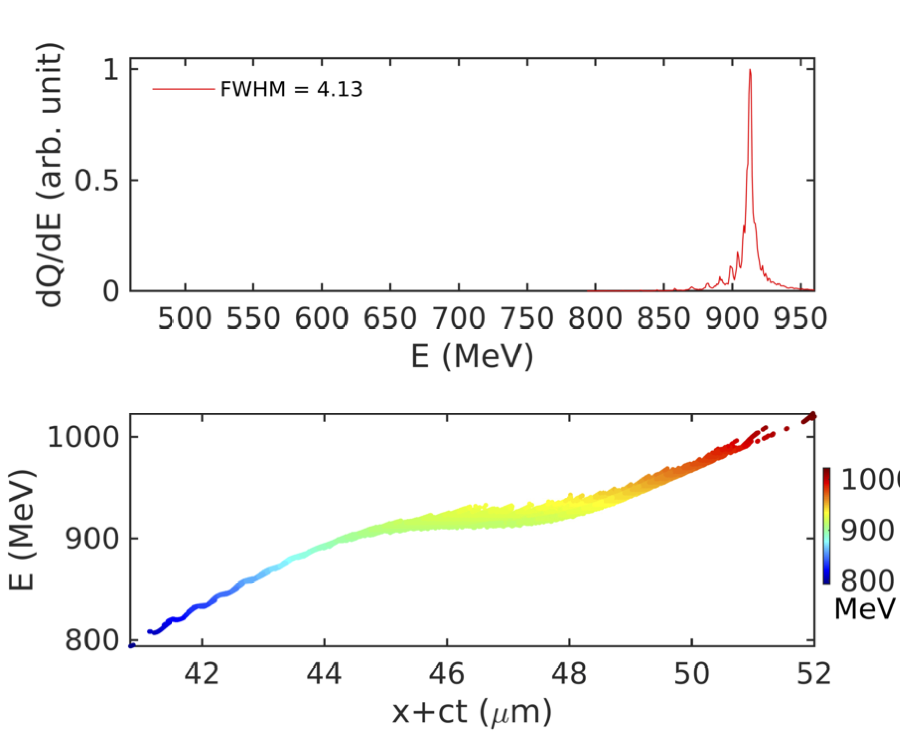}
\caption{}\label{fig:fig_f}
\end{subfigure}
%
\begin{minipage}[t]{.8\textwidth}
\vspace{20pt}
\caption{\small Dynamic evolution of the energy spectrum and longitudinal phase-space of an initially negative energy chirped electron beam with an initial enegy of 500 MeV.}
\end{minipage}

\end{figure}


\begin{figure}
\centering

\begin{subfigure}[t]{.45\textwidth}
\centering
\includegraphics[width=\linewidth]{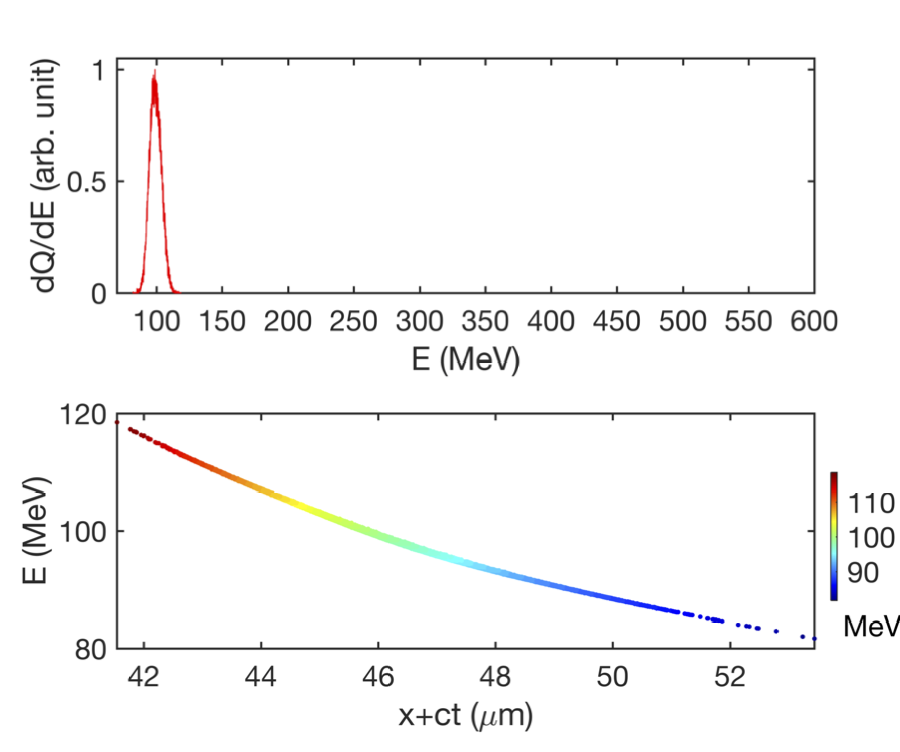}
\caption{}\label{fig:fig_a}
\end{subfigure}
\hskip 20pt
\begin{subfigure}[t]{.45\textwidth}
\centering
\includegraphics[width=\linewidth]{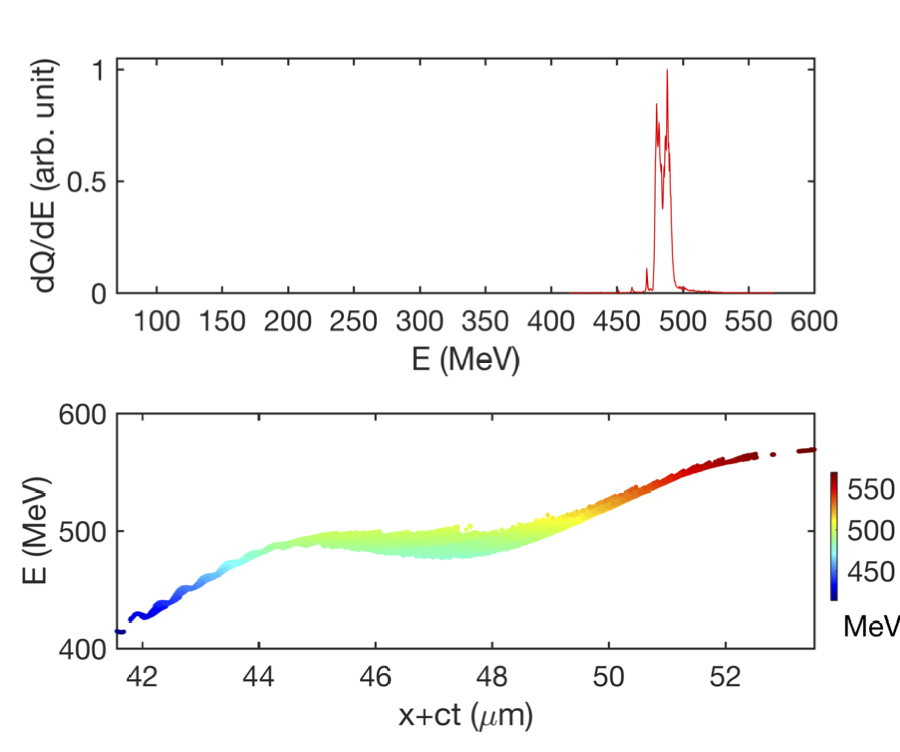}
\caption{}\label{fig:fig_b}
\end{subfigure}

\begin{minipage}[t]{.8\textwidth}
\vspace{20pt}
\caption{\small Dynamic evolution of the energy spectrum and longitudinal phase-space of an initially negative energy chirped electron beam with an initial enegy of 100 MeV and energy spread of $10 \%$.}
\end{minipage}

\end{figure}


\end{document}